\documentclass[11pt, letterpaper]{article}

\usepackage{graphicx}
\usepackage{subcaption}
\usepackage{booktabs}
\usepackage{tabularray}
\UseTblrLibrary{booktabs, siunitx}

\usepackage{amsmath}
\usepackage{amsthm}
\usepackage[thmtools-compat]{keytheorems}   
\usepackage{amssymb}
\usepackage{mathrsfs}
\usepackage{mathtools}
\usepackage{bm}
\usepackage{nicefrac}
\usepackage{dsfont}

\usepackage{siunitx}
\usepackage[dvipsnames]{xcolor}

\usepackage{newfloat}
\DeclareFloatingEnvironment[fileext=lop, placement=htbp, name=LP]{linprog}
\DeclareFloatingEnvironment[fileext=iop, placement=htbp, name=IP]{intprog}

\usepackage{algorithm}
\usepackage{algpseudocode}

\usepackage{xurl}
\usepackage[backref=page,colorlinks=true]{hyperref}
\hypersetup{
    linkcolor=blue,
    citecolor=Plum,
    urlcolor=blue,
}
\usepackage[capitalise,nameinlink]{cleveref}

\usepackage[shortlabels]{enumitem}
\setlist[enumerate]{noitemsep,topsep=6pt,parsep=0pt,partopsep=0pt}
\setlist[itemize]{noitemsep,topsep=0pt,parsep=0pt,partopsep=0pt}

\usepackage{xspace}
\usepackage{fontawesome5}
\usepackage{soul}

\usepackage{tikz}
\usetikzlibrary{
    positioning,
    shapes,
    arrows.meta,
    calc,
    patterns.meta,
    patterns,
    decorations.pathreplacing,
    backgrounds,
    tikzmark,
    fit,
}

\colorlet{mmcclient}{blue!65!black}
\colorlet{mmcclientfill}{blue!8}
\colorlet{mmcfacility}{green!45!black}
\colorlet{mmcfacilityfill}{green!8}
\colorlet{mmcsteiner}{black!65}
\colorlet{mmcsteinerfill}{black!4}

\colorlet{mmcedge}{black!70}
\colorlet{mmcedgefaint}{black!45}
\colorlet{mmcgroup}{black!45}
\colorlet{mmctext}{black!70}

\definecolor{mmcaccentone}{RGB}{145, 70, 150}
\definecolor{mmcaccenttwo}{RGB}{214, 137, 16}

\newlength{\mmclinewidth}
\newlength{\mmcthicklinewidth}
\tikzset{
    mmc/client/.style={
            circle,
            draw=mmcclient,
            fill=mmcclientfill,
            line width=\mmclinewidth,
            minimum size=8mm,
            inner sep=1pt,
            font=\small
        },
    mmc/center/.style={
            rectangle,
            rounded corners=2pt,
            draw=mmcfacility,
            fill=mmcfacilityfill,
            line width=\mmclinewidth,
            minimum width=10mm,
            minimum height=7mm,
            inner sep=2pt,
            font=\small
        },
    mmc/steiner/.style={
            circle,
            draw=mmcsteiner,
            fill=mmcsteinerfill,
            line width=\mmclinewidth,
            minimum size=8mm,
            inner sep=1pt,
            font=\small
        },
    mmc/edge/.style={
            draw=mmcedge,
            line width=\mmclinewidth
        },
    mmc/edge light/.style={
            draw=mmcedgefaint,
            line width=\mmclinewidth
        },
    mmc/leader/.style={
            draw=mmcgroup,
            line width=0.6pt,
            densely dashed
        },
    mmc/arrow/.style={
            mmc/edge,
                -{Latex[length=2.4mm, width=1.8mm]}
        },
    mmc/accent one/.style={
            draw=mmcaccentone,
            line width=\mmcthicklinewidth,
            -{Latex[length=2mm, width=1.6mm]}
        },
    mmc/accent two/.style={
            draw=mmcaccenttwo,
            line width=\mmcthicklinewidth,
            -{Latex[length=2mm, width=1.6mm]}
        },
    mmc/edge label/.style={
            font=\scriptsize,
            fill=white,
            inner sep=2pt
        },
    mmc/note/.style={
            font=\scriptsize,
            text=mmctext
        },
    mmc/length/.style={
            font=\scriptsize
        },
    mmc/divider/.style={
            draw=mmcgroup!60,
            line width=0.8pt
        },
    mmc/group/.style={
            draw=mmcgroup,
            dashed,
            line width=0.8pt,
            rounded corners=4pt
        },
    mmc/stage/.style={
            draw=mmcclient,
            fill=mmcclientfill,
            rounded corners=2pt,
            line width=\mmclinewidth,
            align=left,
            text width=84mm,
            inner xsep=6pt,
            inner ysep=5pt,
            font=\small
        },
    mmc/stage ours/.style={
            mmc/stage,
            draw=mmcfacility,
            fill=mmcfacilityfill
        },
    mmc/stage arrow/.style={
            mmc/arrow,
            line width=\mmcthicklinewidth
        },
    mmc/stage note/.style={
            font=\scriptsize,
            text=mmctext,
            anchor=west,
            inner sep=2pt
        },
}

\DeclarePairedDelimiter{\abs}{\lvert}{\rvert}
\DeclarePairedDelimiter{\norm}{\lVert}{\rVert}

\DeclarePairedDelimiter{\set}{\{}{\}}

\DeclarePairedDelimiter{\ceil}{\lceil}{\rceil}
\DeclarePairedDelimiter{\paren}{(}{)}
\DeclarePairedDelimiter{\bracket}{[}{]}

\DeclareMathOperator{\diag}{diag}

\DeclareMathOperator*{\Expectation}{\mathbb{E}}
\newcommand*{\EE}[2][]{\Expectation_{#1}\bracket*{#2}}

\newcommand*{\bigomega}[1]{\Omega\paren*{#1}}

\newcommand*{\ZZ}{\mathbb{Z}}

\newcommand*{\RR}{\mathbb{R}}

\newcommand*{\calD}{\mathcal{D}}
\newcommand*{\calE}{\mathcal{E}}

\newcommand*{\calP}{\mathcal{P}}

\NewDocumentCommand{\cost}{o g}{%
    \ensuremath{\mathsf{cost}%
        \IfValueT{#1}{_{#1}}%
        \IfValueT{#2}{\!\left(#2\right)}%
    }\xspace%
}

\declaretheorem[name=Theorem, style=plain, numberwithin=section]{theorem}
\declaretheorem[name=Lemma, style=plain, sibling=theorem]{lemma}

\declaretheorem[name=Corollary, style=plain, sibling=theorem]{corollary}
\declaretheorem[name=Definition, style=definition, sibling=theorem]{definition}
\declaretheorem[name=Remark, style=remark, sibling=theorem]{remark}

\DeclareMathOperator{\Lip}{Lip}

\DeclareMathOperator{\conv}{conv}
\DeclareMathOperator{\Ham}{Ham}

\usepackage[margin=1in]{geometry}
\usepackage[T1]{fontenc}
\usepackage[final]{microtype}
\usepackage[numbers,square,sort&compress]{natbib}

\newcommand{\paragraphname}[2][0.75em]{\vspace{#1}\noindent\textbf{#2.}}

\newtheorem{claim}[theorem]{Claim}
\newtheorem{fact}[theorem]{Fact}
\crefname{claim}{Claim}{Claims}
\crefname{fact}{Fact}{Facts}

\crefname{claim}{Claim}{Claims}
\Crefname{claim}{Claim}{Claims}
\crefname{fact}{Fact}{Facts}
\Crefname{fact}{Fact}{Facts}

\DeclareMathOperator{\dist}{dist}

\DeclareMathOperator{\minext}{min-ext}
\newcommand{\lipext}{\mathsf{e}}
\newcommand{\R}{\mathbb R}

\newcommand{\Id}{\mathrm{Id}}
\newcommand{\one}{\mathbf 1}
\newcommand{\ip}[2]{\left\langle #1,#2\right\rangle}
\newcommand{\dd}{\,\mathrm d}

\usepackage{color-edits}
\addauthor[DC]{dc}{magenta!99!black}
\addauthor[AV]{av}{red!99!black}
\addauthor[YM]{ym}{blue!99!black}
\addauthor[SM]{sm}{green!99!black}

\title{Bourgain's Theorem for a Union of Euclidean Metric Spaces}

\usepackage{authblk}
\author[1]{Dimitar Chakarov\thanks{\texttt{chakarov@ttic.edu}}}
\author[2]{Sepideh Mahabadi\thanks{\texttt{smahabadi@microsoft.com}}}
\author[1]{Yury Makarychev\thanks{\texttt{yury@ttic.edu}. Supported in part by NSF award ECCS-2216899.}}
\author[3]{Ali Vakilian\thanks{\texttt{a.vakilian@rutgers.edu}. Supported in part by NSF CAREER award CAIF-2638918.}}
\affil[1]{Toyota Technological Institute at Chicago}
\affil[2]{Microsoft Research}
\affil[3]{Rutgers Business School}

\date{}

\allowdisplaybreaks
\begin{document}

\maketitle

\begin{abstract}
    We study the distortion of embedding a union of $k$ Euclidean sets into Euclidean space.
We show that a union of $k$ Euclidean sets can be embedded with distortion at most $O(k^{1.128} \log k)$, independent of the number of points.
This is an exponential improvement over the previous best known upper bound of $2^{O(k)}$ obtained by iterating the result of Makarychev and Makarychev (2016) for the special case $k = 2$.
Surprisingly, with the additional assumption that each Euclidean set remains Euclidean when we adjoin any other two points, we can obtain an asymptotically tight distortion of $O(\log k)$, thus, providing a generalization of Bourgain's Theorem (1985).
Along the way, we also show that the distortion of the union can be upper bounded by $O(\log k)$ times the maximum of the Lipschitz extension constants of individual Euclidean sets.

\end{abstract}

\section{Introduction}
\label{sec:introduction}

We consider the problem of embedding a union of Euclidean metric spaces into Euclidean space with low distortion.
Formally, let $(X, d)$ be a metric space, such that $X$ is a union of $k$ sets $U_1, \dots, U_k$ and each $U_i$ can be embedded isometrically into Euclidean space.
Our goal is to find the smallest distortion $D_k$, such that the union $U_1 \cup \dots \cup U_k$ embeds into $\ell_2$ with distortion~$D_k$.
When every $U_i$ is a single point this is exactly the setting of Bourgain's Theorem~\cite{Bou85}, which states that every $n$-point metric space can be embedded into Euclidean space with distortion $O(\log n)$.
More generally, our question belongs to the study of local-global properties of metric embeddings initiated by Arora, Lovasz, Newman, Rabani, Rabinovich, and Vempala~\cite{ALNRRV12} (see also \cite{CMM10}), and asks whether the distortion of $X$ is governed only by $k$.

The special case $k = 2$ was asked by Assaf Naor and settled by Makarychev and Makarychev \cite{union16makarychev}, who showed that a union of two Euclidean sets embeds into $\ell_2$ with distortion at most $8.93$.
They then posed the problem for general $k$ as an open question~\cite[Question~6]{union16makarychev}.
On one hand, iterating their result gives an upper bound of $D_k \leq 2^{O(k)}$.
On the other hand, when each $U_i$ is a single point we have the classical lower bound of $D_k = \Omega(\log k)$ for Bourgain's Theorem~\cite{LLR95, AR98}.

\smallskip
In this work, we show that a union of $k$ Euclidean sets can be embedded into Euclidean space with distortion $D_k = O(k^{1.128} \log k)$, an exponential improvement over the previous bound, with no dependence on the number of points in $X$ or any~$U_i$.
Given the additional property that every~$U_i$ stays Euclidean after adjoining any other two points from $X$, we can further establish that the distortion of embedding $X$ is $O(\log k)$, thus, matching the lower bound.
Finally, if each $U_i$ remains Euclidean after adjoining any one other point from $X$, then we can obtain distortion of $O(k^{0.895} \log k)$.

Our approach contains two reductions that may be of independent interest.
First, the distortion of a union of Euclidean sets is determined, up to an $O(\log k)$ factor, by how well Lipschitz maps on the individual sets extend to the whole space.
Second, bounding these Lipschitz extension constants is equivalent to constructing good Euclidean metric sparsifiers.

\subsection{Model and Main Results}

\paragraphname[0em]{Distortion} We say that a metric space $(X, d)$ embeds in $\ell_2$ with \emph{distortion} $D$ if there is a function $f : X \to \ell_2$, such that for every $x, y \in X$:
\begin{equation}
    d(x, y) \leq \norm{f(x) - f(y)}_2 \leq D \cdot d(x, y).
\end{equation}
Then, the \emph{distortion $c_2(X)$ of $(X, d)$} is the smallest such scalar $D$.

\paragraphname{Euclidean Classes}
In a metric space $(X, d)$, a subset $V \subseteq X$ is \emph{Euclidean in $X$} if it embeds isometrically into $\ell_2$.
Moreover, we say a subset $V \subseteq X$ is \emph{$h$-strongly Euclidean in $X$} if, for every $h$-element subset $S \subseteq X$, the set $V \cup S$ embeds isometrically into $\ell_2$.
That is, an $h$-strongly Euclidean set is a Euclidean set that remains Euclidean if we add $h$ arbitrarily chosen points.

The parameter $h$ interpolates between the condition that each $U_i$ is Euclidean ($h = 0$) and that~$X$ is Euclidean itself ($h = \abs{X}$).
In the spirit of local-to-global phenomena~\cite{ALNRRV12, CMM10}, one can ask how large $h$ has to be before the distortion of $X$ becomes the optimal $\Theta(\log k)$.
Our first main result shows that $h = 2$ is already sufficient.

\begin{theorem}\label{thm:main}
    Let $(X, d)$ be a metric space, such that $X = U_1 \cup \cdots \cup U_k$ and each $U_i$ is 2-strongly Euclidean in $X$.
    Then $X$ embeds into $\ell_2$ with distortion $O(\log k)$.
\end{theorem}

Observe that when each $U_i$ consists of a single point, \cref{thm:main} recovers Bourgain's Theorem~\citep{Bou85}.
Because an $h$-strongly Euclidean set is $2$-strongly Euclidean for every $h \geq 2$, the same bound holds for unions of $h$-strongly Euclidean sets with $h \geq 3$.
Therefore, what remains are, indeed, only unions of Euclidean sets and unions of $1$-strongly Euclidean sets.
In the sections that follow we first develop the proof of \cref{thm:main}, and later adapt what is useful to the Euclidean and 1-strongly Euclidean settings.

In the general case of a union of $k$ Euclidean sets we obtain the following subquadratic upper bound on the distortion.

\begin{theorem}\label{thm:euclidean}
    Let $(X, d)$ be a metric space, such that $X = U_1 \cup \cdots \cup U_k$ and each $U_i$ is Euclidean in $X$.
    Then $X$ embeds into $\ell_2$ with distortion $O(k^{1.128} \log k)$.
    Moreover, for the special case $k = 2$ we get an improved upper bound of $c_2(X) < 3.997$.
\end{theorem}

For the case of 1-strongly Euclidean sets, we get the following sublinear distortion.

\begin{theorem}\label{thm:one-strongly-euclidean}
    Let $(X, d)$ be a metric space, such that $X = U_1 \cup \cdots \cup U_k$ and each $U_i$ is 1-strongly Euclidean in $X$.
    Then $X$ embeds into $\ell_2$ with distortion $O(k^{0.895} \log k)$.
\end{theorem}

\subsection{Technical Overview}

\paragraphname[0em]{Union of 2-strongly Euclidean sets}
In \cref{sec:two-strongly-euclidean}, we prove \cref{thm:main} in two steps:
\begin{enumerate}
    \item For each $U_i$ we construct a \emph{$D$-terminal embedding}, a $1$-Lipschitz map $\varphi_i : X \to \ell_2$ that contracts distances between $U_i$ and every point of $X$ by at most a factor of $D$.
          This is where we use the $2$-strongly Euclidean property of the sets $U_i$.
    \item We combine the $k$ terminal embeddings into a single embedding for $X$.
          In the combined embedding, each point uses only the terminal embedding of the set closest to it, and we sample subcollections of the sets as in the algorithmic proof of Bourgain's Theorem~\cite{LLR95}.
\end{enumerate}
Overall, this yields an embedding of $X$ into $\ell_2$ with distortion $O(\log k)$.
We note that our notion of terminal embedding is slightly different from the one considered by \cite{EFN17} because they do not require the maps to be Lipschitz on the whole space, as we do here.

\paragraphname{From distortion to Lipschitz extension constants}
Only the first step above depends on the sets being $2$-strongly Euclidean, while combining terminal embeddings works for any union of Euclidean sets as long as each set admits a terminal embedding.
In \cref{sec:lipschitz-extendability-approximates-distortion}, we show that any $L$-Lipschitz map $X \to \ell_2$ that does not contract distances on $U_i$ can be augmented into an $O(L)$-terminal embedding for $U_i$~(\cref{lem:terminal-from-extension}).
So now the question becomes how well an isometry of $U_i$ extends to a Lipschitz map on all of $X$.
This is captured by the \emph{Lipschitz extension constant} $\lipext(U_i)$, the least $\Lambda \geq 1$ such that every Lipschitz map $f : (U_i, d) \to \ell_2$ extends to a Lipschitz map $F : (X, d) \to \ell_2$ with $\Lip(F) \leq \Lambda \Lip(f)$.
Indeed, since $U_i$ is Euclidean, by Kirszbraun's Theorem~\cite{kirszbraun1934zusammenziehende} it suffices to extend an isometry of $U_i$ in order to bound $\lipext(U_i)$.
The formal result connecting distortion to Lipschitz extension constants is the following.



\begin{theorem}\label{thm:extendability}
    Let $X = U_1 \cup \cdots \cup U_k$ with each $U_i$ Euclidean in $X$, and let $\lipext_{\max} = \max_i \lipext(U_i)$.
    Then
    \begin{equation*}
        \lipext_{\max} \leq c_2(X) \leq O(\lipext_{\max} \log (k + 1)).
    \end{equation*}
\end{theorem}

For instance, for a union of $2$-strongly Euclidean sets, we can interpret the analysis of \cref{thm:main} as showing that $\lipext_{\max} = \Theta(1)$.


\paragraphname{Lipschitz extension constants for Euclidean and 1-strongly Euclidean sets}
In \cref{sec:euclidean-unions}, we bound the Lipschitz extension constants using \emph{metric sparsifiers}~\cite{MM16}.
Given a weighted graph $G$ on $X$ and a set of terminals $A \subseteq X$, a metric sparsifier of $G$ is a weighted graph $H$ on $A$ alone, such that for every metric $\rho$ on $A$ the cost of $H$ on~$\rho$ approximates, up to a factor $Q$ called the \emph{quality}, the least cost of $G$ over all extensions of $\rho$ to $X$.
We take $A = U_i$ and measure cost by the quadratic energy $\calE_G(d^2) = \sum_{x < y} g_{xy} d(x, y)^2$, while only allowing extensions that are Euclidean on every $U_j$.
In \cref{sec:euclidean-sparsifiers}, we show that in this setting the best sparsifier quality for every $G$ is exactly the square of the Lipschitz extension constant, where both quantities are taken uniformly over all metrics on $X$ that are Euclidean on every $U_j$ (\cref{thm:euclidean-equivalence}).
Bounding $\lipext_{\max}$ thus reduces to constructing a good metric sparsifier.

The following upper bounds, combined with \cref{thm:extendability}, give \cref{thm:euclidean,thm:one-strongly-euclidean} for finite $X$.
We take care of the finiteness assumption in \cref{app:finite-reduction}.

\begin{lemma}\label{lemma:lipschitz-constants}
    Let $(X, d)$ be a finite metric space, such that $X = U_1 \cup \cdots \cup U_k$.
    If each $U_i$ is Euclidean in $X$, then $\lipext_{\max} = O(k^{1.128})$.
    If each~$U_i$ is $1$-strongly Euclidean in $X$, then $\lipext_{\max} = O(k^{0.895})$.
\end{lemma}

In the construction for \cref{lemma:lipschitz-constants} we bound the quality of the canonical sparsifier for $U_i$ by adding the other sets back one at a time.
We show that adding one Euclidean set costs a constant factor in the Lipschitz constant, and averaging the extensions over the set added last decreases the factor at the step with $r$ sets down to $1 + \eta / r$ (\cref{lem:improved-averaging}).
The product of these factors is $O(k^{\eta})$, where the exponent $\eta$ is determined by the best trade-off between the growth on pairs of points inside the added set and on pairs crossing it, giving $\eta_0 = 1.1275\ldots < 1.128$ for Euclidean sets and $\eta_1 = 0.8941\ldots < 0.895$ for $1$-strongly Euclidean sets (\cref{sec:improved-extension-bounds}).
We leave the improved upper bound for the case $k = 2$ to \cref{app:two-pieces}.


\paragraphname{Lower bounds on Lipschitz extension constants}
\cref{thm:extendability} bounds the distortion with the maximal Lipschitz extension constant $\lipext_{\max}$ up to a factor of $O(\log k)$.
Meanwhile, the analysis of \cref{thm:main} shows that $\lipext_{\max}$ is an absolute constant for unions of $2$-strongly Euclidean sets.
What happens for unions of Euclidean sets, or of $1$-strongly Euclidean sets?
We find that the extension constant can be $\Omega(\sqrt{\log k})$ for some unions of Euclidean sets, and $\Omega(\sqrt[4]{\log k})$ for 1-strongly Euclidean.
The lower bounds we obtain here do not improve on the general $\Omega(\log k)$ lower bound for distortion.
They do show, however, that the reduction to Lipschitz extensions cannot by itself give $O(\log k)$ for unions of Euclidean sets.

\begin{lemma}\label{lem:lipschitz-lower-bounds}
    There are unions of $k$ Euclidean sets with $\lipext_{\max} = \Omega(\sqrt{\log k})$, and unions of $k$ $1$-strongly Euclidean sets with $\lipext_{\max} = \Omega(\sqrt[4]{\log k})$.
\end{lemma}

In \cref{sec:euclidean-lower-bound} for unions of Euclidean sets, we build $X$ from a graph with $k$ vertices and girth $\Theta(\log k)$, under a specially-designed norm on $\mathbb{R}^E$, so that the cycle space of the graph is Euclidean, with one set per vertex.
Then extending an isometry of one set is the same as a linear projection onto the cycle space.
By a theorem of Lindenstrauss~\cite{lindenstrauss1964nonlinear}, its Lipschitz constant is controlled by the relative projection constant of the cycle space.
In \cref{sec:one-strongly-euclidean-lower-bound}, for unions of $1$-strongly Euclidean sets we use a variant of the twisted cube of Naor and Rabani~\cite{NR17}.

\paragraphname{From isometric embeddings to constant-distortion embeddings}
Finally, in \cref{sec:constant-distortion}, we relax the assumption that each $U_i$ is isometric to a subset of $\ell_2$ and only require that it embeds with constant distortion $C_i$.
Similarly to the isometric setting, the distortion is naively upper bounded by $2^{O(k)} \prod_{i = 1}^k C_i$, and the second term is itself exponential in $k$ when all $C_i$ equal a common constant.
Fortunately, we can do much better than that.

\begin{theorem}\label{thm:euclidean-constant-distortion}
    Let $(X, d)$ be a metric space, such that $X = U_1 \cup \cdots \cup U_k$ and each $U_i$ embeds in $\ell_2$ with distortion $C_i \geq 1$, and let $C_{\max} = \max_i C_i$.
    Then $X$ embeds into $\ell_2$ with distortion $O(C_{\max} \, k^{\eta(C_{\max})} \log k)$, where $\eta(c) \leq c - 1 + \pi / c$ for every $c \geq 1$ and $\eta(1) = \eta_0$.
\end{theorem}

In the isometric setting the 2-strongly Euclidean property is what gave the tight $O(\log k)$ distortion in \cref{thm:main}.
We might hope for a similar effect here, by assuming that every $U_i \cup \{x, y\}$ embeds with constant distortion.
However, by \cite[Theorem~1.1]{union16makarychev}, if $U_i$ embeds with distortion~$C$ then $U_i \cup \{x, y\}$ embeds into $\ell_2$ with distortion $O(C)$ because $\{x, y\}$ is itself Euclidean.
Therefore, up to constants, this condition already follows from the assumption that each $U_i$ embeds with constant distortion.
As a result, adjoining two points provides no meaningful improvement, unlike in the isometric setting.

\subsection{Related Work}

\paragraphname[0em]{Bourgain's Theorem}
While the initial result of Bourgain~\cite{Bou85} was existential, Linial, London and Rabinovich~\cite{LLR95} gave an algorithmic construction and initiated the use of low-distortion embeddings in approximation algorithms, with early example applications being~\cite{LLR95,AR98,Fei00}.
Beyond approximation algorithms, Bourgain's Theorem has been used for Internet latency estimation~\cite{KSW09}, differential privacy~\cite{huang14differentialprivacy}, and designing position-aware embeddings for graph neural networks~\cite{YYL19}.
We refer the reader to~\cite{Ind01, Lin02} for surveys on low-distortion embeddings.


\paragraphname{Terminal embeddings}
Elkin, Filtser and Neiman~\cite{EFN17} considered a different generalization of Bourgain's Theorem.
Given a set of terminals $T \subseteq X$, a terminal embedding approximately preserves the distance from every point of $X$ to every terminal, and they obtained distortion $O(\log \abs{T})$.
When $X$ is itself Euclidean, Mahabadi, Makarychev, Makarychev and Razenshteyn~\cite{MMMR18} showed that terminal embeddings can be obtained by extending an embedding of the terminals to the whole space with distortion $1 + \varepsilon$ and $O(\varepsilon^{-4} \log \abs{T})$ dimensions. Later Narayanan and Nelson~\cite{NN19} improved the number of dimensions to the tight $O(\varepsilon^{-2} \log \abs{T})$, and Cherapanamjeri and Nelson~\cite{CN24} showed that such embeddings can be computed in time sublinear in $\abs{T}$.


\section{Union of 2-strongly Euclidean Sets}
\label{sec:two-strongly-euclidean}

In this section we prove \cref{thm:main}.
We show how to construct terminal embeddings for 2-strongly Euclidean sets, and then combine those using a procedure inspired by Bourgain's Theorem.

\subsection{Preliminaries}

\paragraphname{Notation}
For $1\le p<\infty$, we write $A\oplus_p B$ for the direct sum with norm $\|(a,b)\|=(\|a\|^p+\|b\|^p)^{1/p}$, and $\bigoplus_{p,\,i=1}^r A_i$ for a finite sum of this kind.
For $n < m$, we will identify $\ell_2^n$ with the subspace of $\ell_2^m$ spanned by the first $n$ coordinates.
For a finite- or infinite-dimensional space $\ell_2^r$, the space $\ell_2^r \oplus_2 \mathbb R$ is the Euclidean or Hilbert space obtained from $\ell_2^r$ by adding an extra coordinate; we will identify $\ell_2^r$ with the subspace of $\ell_2^r \oplus_2 \mathbb R$ where the extra coordinate is 0.

\paragraphname{Two-sided Kirszbraun theorem for combining terminal embeddings}
This lemma allows us to smoothly combine terminal embeddings for different sets $U_i$.

\begin{lemma}[{\cite[Cor.~14]{BMMM21}}]\label{cor:collapse}
    Let $S$ be a nonempty subset of $\ell_2^m$, where $m$ may be finite or infinite.
    Define $R(x)=\dist(x,S)$.
    There exists a map $\psi: \ell_2^m\to\ell_2$ such that:
    \begin{enumerate}
        \item $\psi(x)=0$ for every $x\in S$;
        \item for every $x,y\in \ell_2^m$,
              \[
                  \|\psi(x)-\psi(y)\| = \Theta(\min\bigl\{\|x-y\|,R(x)+R(y)\bigr\}).
              \]
    \end{enumerate}
    The hidden constants in the $\Theta$ notation are absolute and do not depend on $S$ and $m$.
\end{lemma}
Fix a finite or infinite $n$.
Let $m = n +1$.
Applying \cref{cor:collapse} to $S = \ell_2^n = \{(u,0): u\in \ell_2^n\} \subset \ell_2^n \oplus_2 \mathbb{R}$, we get the following corollary.
\begin{corollary}\label{cor:psi}
    For every $n$, there exists a map $\psi: \ell_2^n \times \mathbb{R} \to \ell_2$ with the following properties:
    \begin{itemize}
        \item $\psi(u,0) = 0$ for all $u\in \ell_2^n$.
        \item $\|\psi(u,a) - \psi(v,b)\| = \Theta(\min(\|u - v\| + |a-b|, a + b))$ for all $u,v\in \ell_2^n$ and $a,b\geq 0$, where the constants in the $\Theta$-notation are absolute.
    \end{itemize}
\end{corollary}
We will construct maps $\psi$ for different values of $n$, but use the same notation $\psi$ for all of them, since the value of $n$ will be clear from context.
Moreover, all these maps can be viewed as restrictions of the map $\psi$ for $n=\infty$ to finite values of $n$.

\paragraphname{Reduction to finite spaces}
For bounding the distortion, it is enough to consider only finite metric spaces.
Let $X = U_1 \cup \cdots \cup U_k$ be as in any of \cref{thm:main,thm:euclidean,thm:one-strongly-euclidean,thm:extendability,thm:euclidean-constant-distortion}.
Since each $U_i$ embeds into $\ell_2$ with finite distortion, $X$ is separable.
Then the compactness argument of \cite[Lemma~2.9, part~I]{union16makarychev} shows that $X$ embeds with distortion $D$ as long as every finite subset of $X$ does.
Moreover, every finite $S \subseteq X$ is covered by the nonempty sets $S \cap U_i$, which inherit the Euclidean property of $U_i$.
Further details are in \cref{app:finite-reduction}.
For the rest of the paper we assume that all metric spaces are finite.

\subsection{Terminal embeddings for 2-strongly Euclidean subsets}
We now show that the 2-strongly Euclidean property implies the existence of $1$-terminal embeddings.

\begin{definition}
    Let $(X,d)$ be a metric space, and let $V\subseteq X$ be a nonempty subset. For $D\ge1$, a map $\varphi:  X\to\ell_2^m$, where $m$ may be finite or infinite, is a $D$-terminal embedding for $V$ if it satisfies the following properties:
    \begin{enumerate}
        \item The map $\varphi$ is $1$-Lipschitz.
        \item The map $\varphi$ contracts distances from points in $V$ by at most a factor $D$: for every
              $x\in V$ and $y\in X$,
              \[
                  \|\varphi(x)-\varphi(y)\|\ge d(x,y)/D.
              \]
              In particular, when $D=1$, $\varphi$ preserves all distances from points in $V$.
    \end{enumerate}
\end{definition}

We use a variant of terminal embeddings introduced by \cite{EFN17}.
In their terminology, we require our $D$-terminal embeddings for $V$ to have terminal distortion at most $D$ and to be $1$-Lipschitz on all of $X$; the latter condition is not required in their original definition.

Let us say that an isometric embedding $\varphi: Z \to \ell_2^r$ (where $r$ is finite or infinite) is \textit{tight} if the closed affine span $L$ of $\varphi(Z)$ coincides with $\ell_2^r$.
If $\varphi$ is not tight, we can transform it to a tight isometric embedding by restricting the target space to $L$.

We will use the following simple geometric fact.
\begin{fact}[Rigidity of $\ell_2$ embeddings]\label{fact}
    Consider a tight isometric embedding $\varphi: Z \to \ell_2^n$. Then every isometric embedding $\varphi'_x$ of $Z \cup \{x\}$ into $\ell_2$ can be transformed into an isometric embedding $\varphi_x$ into $\ell_2^n \oplus_2 \mathbb{R}$ such that
    \begin{itemize}
        \item $\varphi_x(z) = \varphi(z) \oplus_2 0$ for $z\in Z$ (that is, $\varphi_x$ is an extension of $\varphi$);
        \item $\varphi_x(x) = u \oplus_2 a$ for some $u\in \ell_2^n$ and $a \geq 0$.
    \end{itemize}
\end{fact}

\begin{figure}
    \centering
    \begin{tikzpicture}[font=\small, line cap=round, line join=round]
  \filldraw[fill=black!3, draw=black!55]
    (-3,-0.8) -- (1.8,-0.8) -- (3,1) -- (-1.8,1) -- cycle;
  \begin{scope}[xshift=0.4cm]
  \path[fill=blue!12, draw=blue!55!black, thick]
    (-2.15,-0.15) .. controls (-2.35,0.15) and (-1.75,0.58) .. (-1.25,0.55)
    .. controls (-0.8,0.75) and (-0.15,0.48) .. (-0.35,0.15)
    .. controls (-0.1,-0.12) and (-0.8,-0.42) .. (-1.2,-0.3)
    .. controls (-1.65,-0.5) and (-2,-0.4) .. cycle;
  \node at (-1.25,0.12) {$\varphi_0(V)$};
  \end{scope}
  \node[anchor=north east] at (1.75,-0.85) {$\ell_2^r$};
  \coordinate (u) at (1,0.05);
  \coordinate (x) at (1,2.3);
  \draw[dashed, thick] (u) -- node[right, midway] {$a$} (x);
  \draw[black!55] (u) -- (1.65,0.05);
  \draw (1.2,0.05) -- (1.2,0.25) -- (1,0.25);
  \fill (u) circle (1.6pt) node[below left] {$u$};
  \fill (x) circle (1.8pt) node[above] {$\varphi(x)=u\oplus a$};
\end{tikzpicture}
    \caption{The terminal embedding places $x$ above the subspace containing $\varphi_0(V)$, with projection $u$ and height $a$.
        The projection $u$ need not belong to $\varphi_0(V)$.
    }
    \label{fig:terminal-embedding}
\end{figure}
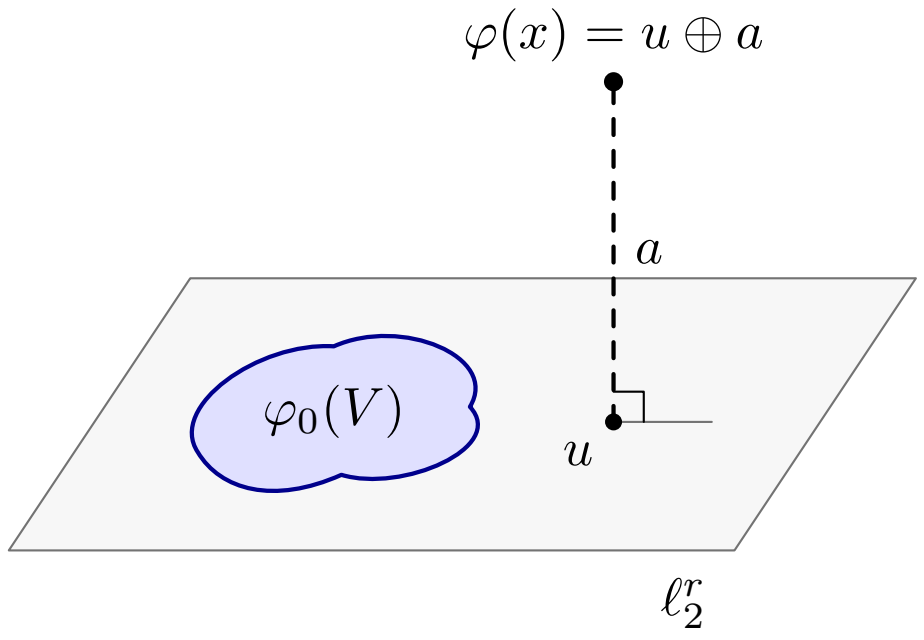

See \cref{fig:terminal-embedding} for the geometry of the terminal embedding.

\begin{claim}\label{terminal}
    If $V$ is a nonempty 2-strongly Euclidean subset of $X$, then there exists a $1$-terminal embedding $\varphi:  X\to\ell_2$ for $V$.
\end{claim}
\begin{proof}
    Since $V$ is 2-strongly Euclidean in $X$, it isometrically embeds into some $\ell_2^r$.
    Let $\varphi_0: V \to \ell_2^r$ be a tight isometric embedding.
    For every $x\in X$, we are also given that $V \cup \{x\}$ isometrically embeds into $\ell_2$.
    By \cref{fact}, there is an embedding $\varphi_x: V \cup \{x\} \to \ell_2^r \oplus_2 \mathbb{R}$ extending $\varphi_0$: $\varphi_x(z) = \varphi_0(z) \oplus_2 0$ for $z\in V$ and $\varphi_x(x) = u \oplus_2 a$ for $u\in \ell_2^r$ and $a\geq 0$.
    We are ready to define $\varphi$:

    \[
        \varphi(x) = \begin{cases}
            \varphi_0(x), \text{ if } x\in V \\
            \varphi_x(x), \text{ if } x\notin V \\
        \end{cases}
    \]
    It is immediate that $\varphi$ preserves distances from points in $V$: for $x\in V$ and $y\in X$,
    \[
        \|\varphi(x) - \varphi(y)\|
        =\|\varphi_y(x) - \varphi_y(y)\|
        = d(x,y).
    \]
    It remains to show that $\varphi$ is 1-Lipschitz. That is,
    \[
        \|\varphi(x) - \varphi(y)\| \leq d(x,y).
    \]
    The claim is trivial if at least one of the points $x$ and $y$ is in $V$ (in this case, we have an equality).
    So let us assume that $x,y\notin V$.

    Using \cref{fact}, extend $\varphi_x$ to a map $\varphi_{xy}: V \cup \{x,y\} \to \ell_2^r \oplus_2 \mathbb{R} \oplus_2 \mathbb{R}$.
    We have
    \begin{align*}
        \varphi_{xy}(x) & = (\varphi_x(x),0) = (\varphi(x),0) = (u, a,0), \\
        \varphi_{xy}(y) & = (v, b \cos \alpha, b \sin \alpha)
    \end{align*}
    for some $u,v\in \ell_2^r$, $a,b \geq 0$ and $\alpha \in [0,2\pi)$. Since both $\varphi_y$ and $\varphi_{xy}$ are isometric embeddings of $V \cup \{y\}$, we further have
    \[
        \varphi(y) = \varphi_y(y) = (v, b).
    \]
    Now we upper bound the distance between $\varphi(x)$ and $\varphi(y)$:
    \[
        \|\varphi(x) - \varphi(y)\|^2 =
        \|u - v\|^2 + a^2 + b^2 - 2ab
        \leq  \|u - v\|^2 + a^2 + b^2 - 2ab\cos \alpha =
        \|\varphi_{xy}(x) - \varphi_{xy}(y)\|^2 = d(x,y)^2,
    \]
    where we used that $\varphi_{xy}$ is an isometry.
\end{proof}

\cref{terminal} immediately gives an embedding with distortion at most $\sqrt{k}$ under the assumptions of \cref{thm:main}.
This guarantee is much weaker than the desired one, but its proof is immediate.
Construct a $1$-terminal embedding $\varphi_i$ for each $U_i$ and let $F = \varphi_1 \oplus_2 \dots \oplus_2 \varphi_k$.
Since all maps $\varphi_i$ are 1-Lipschitz, $\Lip(F) \leq \sqrt{k}$.
On the other hand, for $x\in U_i$ and $y\in X$, we have
\[
    \|F(x) - F(y)\| \geq \|\varphi_i(x) - \varphi_i(y)\|
    = d(x,y).
\]
Therefore, the distortion is at most $\sqrt{k}$.

The construction above concatenates terminal embeddings for the individual sets $U_i$.
Its analogue for a $k$-point metric space concatenates the distance functions to individual points.
For a metric space $X= \{x_1, \dots, x_k\}$, there is a simple embedding $F(y)= (d(x_1,y), d(x_2, y), \dots, (d(x_k,y)))$ of $X$ into $\ell_2^k$ with distortion $\sqrt{k}$.
Bourgain's proof improves on this latter construction: to get an asymptotically optimal distortion $O(\log k)$, we concatenate distance functions to different subsets $A$ of $X$.
Informally, the advantage of this approach is that we have much tighter control over the Lipschitz constant of the obtained map.

Now our goal is to generalize the proof of Bourgain's theorem to our setting.
Our next step is to define a map for a subcollection of $U_1, \dots, U_k$.

\subsection{Combining terminal embeddings}
In this section, we construct a map $f$ for a given collection of nonempty subsets $V_1,\dots, V_r$ of $X$.
We do not assume that $V_1, \dots, V_r$ cover $X$.
In the proof of \cref{thm:main}, we will apply this construction to a randomly chosen subcollection of $U_1, \dots, U_k$.

Let $D\ge1$, and suppose that each $V_i$ has a $D$-terminal embedding $\varphi_i$.
Under the assumptions of \cref{thm:main}, \cref{terminal} gives such embeddings with $D=1$.

To avoid the blow-up in the Lipschitz constant caused by simply concatenating the maps $\varphi_i$, we aim to ensure that $f$ is $C$-Lipschitz for some absolute constant $C$.

Let $N_i$ be the Voronoi cell of $V_i$, defined by
\[
    N_i = \{x: d(x, V_i) < d(x, V_j) \text{ for all }j\neq i\}.
\]
In addition to this Lipschitz condition, we seek a useful lower bound on the distance between the images of $x \in N_1\cup \dots \cup N_r$ and $y\in X$ under $f$.

\paragraph{Construction idea} The key idea is to use the map $\varphi_i$ within each cell $N_i$, avoiding the $\sqrt{k}$ Lipschitz constant blow-up.
However, this idea needs a refinement, because as-is the map will have a huge Lipschitz constant near the ``boundary'' between different Voronoi cells.
To avoid this, we gradually reduce the map to zero at the boundary.
To formalize this idea, we introduce the following parameters.

Let
\begin{align*}
    R_i(x) & = d(x, V_i) \\
    \hat R_i(x) & = \min_{j:j\neq i} d(x, V_j) \\
    \Delta_i(x) & =
    \begin{cases}
        \frac{\hat R_i(x) - R_i(x)}{2} \text{for } x\in N_i \\
        0, \text{otherwise}
    \end{cases}
\end{align*}
We let $\hat{R}_i(x) = \Delta_i(x) = \infty$, if $i = r = 1$ (there is only one set $V_1$).
By definition, $R_i(x) < \hat R_i (x)$ inside $N_i$ and thus $\Delta_i(x) \geq 0$.
Note that the distance to a fixed nonempty set is a 1-Lipschitz function and therefore $R_i$ is 1-Lipschitz.
If $r > 1$, then $\hat R_i$ is also 1-Lipschitz and thus $\hat R_i(x) - R_i(x)$ is 2-Lipschitz.
Therefore, $\Delta_i(x) = \frac{\max(\hat R_i(x) - R_i(x),0)}{2}$ is 1-Lipschitz.
Informally, $\Delta_i(x)$ serves as a proxy for the distance from $x\in N_i$ to the boundary of $N_i$, since the open ball of radius $\Delta_i(x)$ around $x\in N_i$ lies entirely within $N_i$.

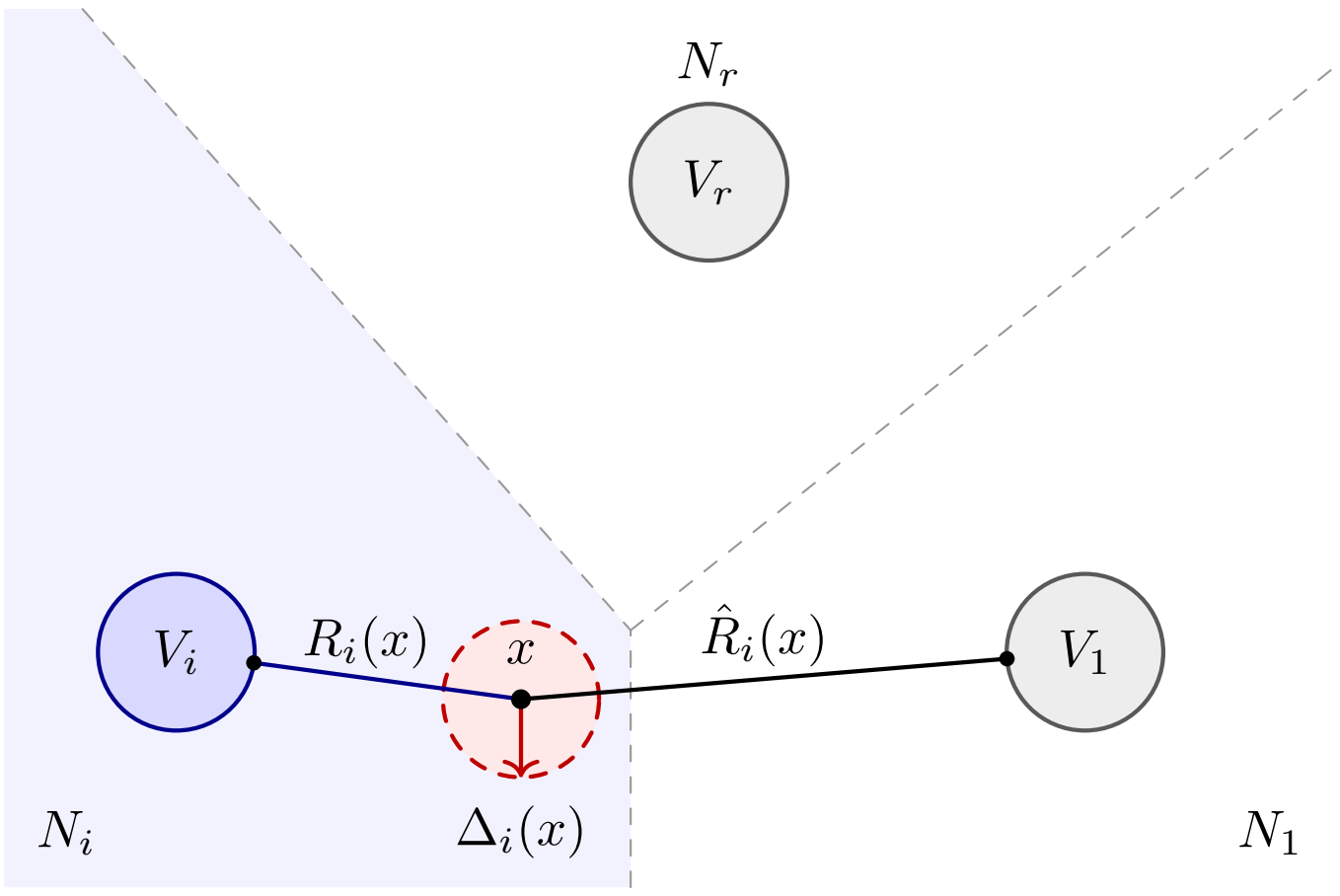
\begin{figure}
    \centering
    \begin{tikzpicture}[font=\small, every node/.style={text=black}, line cap=round, line join=round, scale=1.05]
  \fill[blue!5] (-4.5,-1.5) -- (-0.5,-1.5) -- (-0.5,0.14)
    -- (-4,4.106667) -- (-4.5,4.106667) -- cycle;
  \draw[black!40, dashed] (-0.5,-1.5) -- (-0.5,0.14) -- (-4,4.106667);
  \draw[black!40, dashed] (-0.5,0.14) -- (4,3.74);
  \coordinate (A) at (-3.4,0);
  \coordinate (B) at (2.4,0);
  \coordinate (C) at (0,3);
  \coordinate (X) at (-1.2,-0.3);
  \pgfmathsetmacro{\deltaradius}{(sqrt(3^2+0.2^2)-sqrt(2^2+0.2^2))/2}
  \filldraw[fill=red!9, draw=red!75!black, thick, dashed]
    (X) circle (\deltaradius);
  \filldraw[fill=blue!15, draw=blue!55!black, thick] (A) circle (0.5);
  \filldraw[fill=black!7, draw=black!65, thick] (B) circle (0.5);
  \filldraw[fill=black!7, draw=black!65, thick] (C) circle (0.5);
  \node at (A) {$V_i$};
  \node at (B) {$V_1$};
  \node at (C) {$V_r$};
  \node[anchor=south west, inner sep=0pt] at (-4.3,-1.3) {$N_i$};
  \node[anchor=south east, inner sep=0pt] at (3.8,-1.3) {$N_1$};
  \node[anchor=north, inner sep=0pt] at (0,3.9) {$N_r$};
  \coordinate (nearA) at ($(A)!0.5cm!(X)$);
  \coordinate (nearB) at ($(B)!0.5cm!(X)$);
  \draw[thick, blue!55!black] (X) -- node[above=1pt, inner sep=1pt, midway, xshift=-4pt] {$R_i(x)$} (nearA);
  \draw[thick] (X) -- node[above=1pt, inner sep=1pt, midway] {$\hat R_i(x)$} (nearB);
  \fill (nearA) circle (1.4pt);
  \fill (nearB) circle (1.4pt);
  \draw[->, thick, red!75!black] (X) -- ++(270:\deltaradius)
    node[below=3pt, inner sep=2pt] {$\Delta_i(x)$};
  \fill (X) circle (1.8pt) node[above=3pt] {$x$};
\end{tikzpicture}
    \caption{A schematic Euclidean illustration of the Voronoi cells (not to scale).
        For $x\in N_i$, $R_i(x)$ is the distance to $V_i$, while $\hat R_i(x)$ is the distance to the nearest set $V_j$ other than $V_i$.
    }
    \label{fig:voronoi-distances}
\end{figure}

\cref{fig:voronoi-distances} illustrates the distances to the sets and the open ball of radius $\Delta_i(x)$ contained in $N_i$.

Now we are ready to define the desired embedding $f$ of $X$:
\[
    f(x) = \bigoplus_{2,\,i=1}^r \psi(\varphi_i(x), \Delta_i(x)),
\]
where, for $i = r= 1$, we use the convention that $\psi(\varphi_i(x), \infty) = \varphi_i(x)$.
Note that in this case, $f(x) = \varphi_1(x)$ is 1-Lipschitz.

\begin{theorem}\label{thm:subcollection}
    The map $f$ satisfies the following properties, with constants independent of $D$.
    \begin{enumerate}
        \item $f$ is $C$-Lipschitz for some absolute constant $C$;
        \item for every $i\in [r]$, $x\in N_i$ and $y\in X$,
              \[
                  \|f(x) - f(y)\| \geq \Omega\left(\min\left(\frac{d(x,y)}{2D}, \hat R_i(x)\right) - R_i(x)\right).
              \]
    \end{enumerate}
\end{theorem}
\begin{proof}
    First, we verify that $f$ is $C$-Lipschitz.
    If $r =1$, $f$ is 1-Lipschitz, so let us assume that $r\geq 2$.
    Since $\varphi_i$ is a $D$-terminal embedding for $V_i$, it is 1-Lipschitz.
    We have observed above that $\Delta_i$ is also 1-Lipschitz.
    By \cref{cor:psi}, item 2, $\|\psi(u,a) - \psi(v,b)\| = O(\|u - v\| + |a-b|)$ and thus $\psi$ is $C'$-Lipschitz for some absolute constant $C'$.
    We conclude that $\psi(\varphi_i(x), \Delta_i(x))$ is $C''$-Lipschitz.
    For every $x,y\in X$, at most one coordinate of $f(x)$ is non-zero and at most one of the coordinates of $f(y)$ is non-zero.
    Therefore, we have that $\|f(x) - f(y)\| \leq \sqrt{2} C'' d(x,y)$.
    That is, $f$ is $(\sqrt{2} C'')$-Lipschitz.

    Now we verify the second property.
    We first show that for every $x, y\in X$, we have
    \[
        \|\varphi_i(x) - \varphi_i(y)\| \geq d(x,y)/D - 2R_i(x).
    \]
    Assume first that $V_i$ is compact.
    Let $x'$ be a closest point to $x$ in $V_i$.
    Then $d(x,x') = R_i(x)$.
    Applying the triangle inequality twice, we get
    \begin{equation}\label{eq:double-triangle}
        \begin{aligned}
            \|\varphi_i(x) - \varphi_i(y)\|
             & \geq \|\varphi_i(x') - \varphi_i(y)\| - \|\varphi_i(x) - \varphi_i(x')\| \\
             & \geq d(x',y)/D - d(x,x')
            \geq d(x,y)/D - 2d(x,x'),
        \end{aligned}
    \end{equation}
    as required.
    Here, we used that $\varphi_i$ is $1$-Lipschitz and contracts distances from points in $V_i$, including $x'$, to all points in $X$ by at most a factor $D$.
    If $V_i$ is not compact, choose a sequence $x'$ in $V_i$ such that $d(x,x') \to R_i(x)$ and apply \eqref{eq:double-triangle}.
    We get the desired inequality in the limit.

    Let $x\in N_i$ and $y\in X$.
    If $r\geq 2$, we have
    \begin{align*}
        \|f(x) - f(y)\| & \geq \|\psi(\varphi_i(x), \Delta_i(x)) - \psi(\varphi_i(y), \Delta_i(y))\| \\
         & \geq \Omega(\min(\|\varphi_i(x) - \varphi_i(y)\| + |\Delta_i(x) - \Delta_i(y)|, \Delta_i(x) + \Delta_i(y))) \\
         & \geq \Omega(\min(\|\varphi_i(x) - \varphi_i(y)\|, \Delta_i(x))) \\
         & \geq \Omega(\min(d(x,y)/D - 2R_i(x), \hat R_i(x) - R_i(x))) \\
         & \geq \Omega(\min(d(x,y)/(2D), \hat R_i(x)) - R_i(x)).
    \end{align*}
    For $r = 1$, we have $f(x) = \varphi_1(x)$ and thus $\|f(x) - f(y)\| = \|\varphi_1(x) - \varphi_1(y)\| \geq d(x,y)/D - 2R_1(x) = 2(\min(d(x,y)/(2D), \infty) - R_1(x))$.
\end{proof}

\subsection{Proof of Theorem~\ref{thm:main}}
We prove the following statement, which gives \cref{thm:main} when $D=1$.

\begin{theorem}\label{thm:composition}
    Let $(X,d)$ be a metric space covered by nonempty sets $U_1,\ldots,U_k$.
    Suppose that every $U_i$ has a $D$-terminal embedding into $\ell_2$.
    Then $X$ embeds into $\ell_2$ with distortion $O(D\log(k+1))$.
\end{theorem}

\begin{proof}[Proof of \cref{thm:composition}]
    We construct a random embedding $F$ of $X$ into $\ell_2$.
    As the space of random vectors $L_2(\Omega, \ell_2)$ is isometrically isomorphic to $\ell_2$, $F$ will yield the desired embedding of $X$ into $\ell_2$.
    Let $\ell = \lfloor \log_2 k\rfloor$.
    Define a random subset $A \subseteq [k]$ as follows.
    \begin{itemize}
        \item Choose $s$ uniformly at random from $\{0,1,\dots, \ell\}$.
        \item For every $x$, include $x$ in $A$ independently with probability $1/2^{s+1}$.
        \item Construct the embedding $f_A$ for the sets $U_i$ with $i\in A$ as in the previous section.
              If $A = \emptyset$, define $f_A$ to be the identically zero map.
              Let $F = f_A$.
    \end{itemize}

    We now analyze the distortion of $F$.
    By \cref{thm:subcollection}, item 1, all maps $f_A$ are $C$-Lipschitz for some absolute constant $C$, so the map $F$ is also $C$-Lipschitz.
    Now, fix two points $x,y\in X$.
    We have
    \[
        \|F(x) - F(y)\|_{L_2}= \sqrt{\mathbb{E}[\|F(x)  - F(y)\|^2]} \geq \mathbb{E}_A\|f_A(x) - f_A(y)\|
    \]
    by Jensen's inequality.
    Let $g_i = d(x, U_i)$ for $i\in [k]$.
    Note $x$ belongs to some $U_i$, and thus one of the $g_i$ is $0$.
    Without loss of generality, we may assume that $g_1 \leq g_2 \leq \dots \leq g_k$.
    Then $g_1 = 0$.
    Further, extend the definition of $g_i$ by letting $g_j = \infty$ for $j > k$.

    Let $s_0$ be the random number chosen in the first step of the construction of $A$.
    Let $\mathcal{E}_s$ be the event that exactly one number in $[2^s]$ is included in $A$ and no numbers from $[2^s+1, 2^{s+1}]$ are included in $A$.
    Conditioned on $s=s_0$, the probability of this event is at least some absolute constant.

    \cref{fig:sampling-event} illustrates the event $\mathcal{E}_s$.
    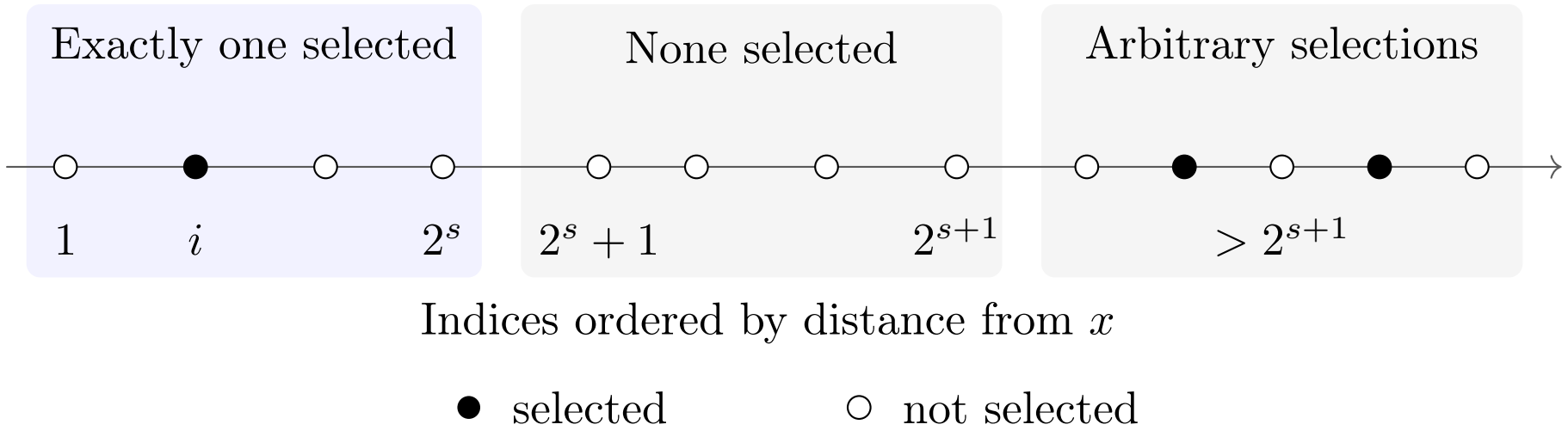
\begin{figure}
        \centering
        \begin{tikzpicture}[font=\small, line cap=round,
  index label/.style={below=5pt, text height=2.2ex, text depth=0.5ex}]
  \fill[blue!5, rounded corners=3pt] (0,-0.85) rectangle (3.5,1.25);
  \fill[black!4, rounded corners=3pt] (3.8,-0.85) rectangle (7.5,1.25);
  \fill[black!4, rounded corners=3pt] (7.8,-0.85) rectangle (11.5,1.25);
  \node at (1.75,0.92) {Exactly one selected};
  \node at (5.65,0.92) {None selected};
  \node at (9.65,0.92) {Arbitrary selections};
  \draw[->, black!65] (-0.15,0) -- (11.8,0);
  \foreach \p in {0.3,1.3,2.3,3.2,4.4,5.15,6.15,7.15,8.15,8.9,9.65,10.4,11.15}
    \draw[fill=white] (\p,0) circle (2.5pt);
  \foreach \p in {1.3,8.9,10.4}
    \fill (\p,0) circle (2.5pt);
  \node[index label] at (0.3,0) {$1$};
  \node[index label] at (1.3,0) {$i$};
  \node[index label] at (3.2,0) {$2^s$};
  \node[index label] at (4.4,0) {$2^s+1$};
  \node[index label] at (7.15,0) {$2^{s+1}$};
  \node[index label] at (9.65,0) {$>2^{s+1}$};
  \node at (5.7,-1.2) {Indices ordered by distance from $x$};
  \fill (3.4,-1.85) circle (2.5pt);
  \node[anchor=west] at (3.6,-1.85) {selected};
  \draw[fill=white] (6.4,-1.85) circle (2.5pt);
  \node[anchor=west] at (6.6,-1.85) {not selected};
\end{tikzpicture}
        \caption{The event $\mathcal{E}_s$: exactly one index in the first block is selected, none in the second, and selections in the remaining block are unrestricted.}
        \label{fig:sampling-event}
    \end{figure}

    Assume $\mathcal{E}_s$ occurs.
    Let $i\in [2^s]$ be the smallest element in $A$.
    Then
    \[
        R_i(x) = d(x, U_i) = g_i \leq g_{2^s}.
    \]
    On the other hand, the second smallest element in $A$, if present, is at least $2^{s + 1} + 1$. Therefore,
    \[
        \hat R_i(x) \geq g_{2^{s + 1} + 1} \geq g_{2^{s + 1}}.
    \]
    If $A$ contains only one element, then the expression on the left-hand side is $\infty$ and the inequality holds trivially.
    Finally, let $t\in \{0,1,\dots,\ell\}$ be the largest integer such that $g_{2^t} \leq d(x,y)/(2D)$.
    The number $t$ is well-defined, since $g_{2^0} = g_1 = 0 \leq d(x,y)/(2D)$ and $g_{2^{\ell+1}} = \infty$.
    For $s\leq t$, if $\min(d(x,y)/(2D), g_{2^{s+1}}) > g_{2^s}$, then $\hat R_i(x) > R_i(x)$ and hence $x\in N_i$; otherwise, the corresponding summand below is zero.
    Applying \cref{thm:subcollection}, item 2, and averaging over the scales, we obtain a telescoping sum:
    \begin{align*}
        \|F(x) - F(y)\|_{L_2} & \geq \mathbb{E}_A\|f_A(x) - f_A(y)\| \\
         & \geq \frac{\Omega(1)}{\ell + 1}\sum_{s=0}^{\ell} \mathbb{E}_A[\|f_A(x) - f_A(y)\| \mid s_0=s,\mathcal{E}_s] \\
         & \geq \frac{\Omega(1)}{\ell + 1}\sum_{s=0}^{t} \left(\min(d(x,y)/(2D), g_{2^{s+1}}) - g_{2^s}\right) \\
         & = \frac{\Omega(1)}{\ell + 1}\left((g_2 - g_1) + (g_4 - g_2) + \dots + (g_{2^{t}} - g_{2^{t- 1}}) + \left(\frac{d(x,y)}{2D} - g_{2^{t}}\right)\right) \\
         & =  \frac{\Omega(d(x,y))}{D(\ell + 1)}.
    \end{align*}
    We conclude that $F$ has distortion $O(D(\ell+1)) = O(D\log(k+1))$, as required.
\end{proof}

\begin{proof}[Proof of \cref{thm:main}]
    By \cref{terminal}, every $U_i$ has a $1$-terminal embedding.
    Apply \cref{thm:composition} with $D=1$.
\end{proof}

\section{Relationship between Distortion and Lipschitz Extension Constants}
\label{sec:lipschitz-extendability-approximates-distortion}

In this section we establish \cref{thm:extendability}, which wants us to show $\lipext_{\max} \leq c_2(X) \leq O(\lipext_{\max} \log (k + 1))$, where $\lipext_{\max} = \max_i \lipext(U_i)$ is the maximal Lipschitz extension constant for any $U_i$.
We first show how to turn a Lipschitz extension into a terminal embedding.
Then \cref{thm:composition} gives an upper bound on the distortion.
The lower bound relies only on Kirszbraun's Theorem~\cite{kirszbraun1934zusammenziehende}.

\subsection{Upper Bound via Terminal Embeddings}

\begin{lemma}[From Lipschitz extension to terminal embedding]\label{lem:terminal-from-extension}
    Let $V$ be a nonempty subset of a finite metric space $(X, d)$, and let $D \geq 1$.
    If a non-contracting embedding $\varphi : V \to \ell_2$ extends to a map $\tilde\varphi : X \to \ell_2$ with $\Lip(\tilde\varphi) \leq D$, then $V$ has a $3 \sqrt{2} D$-terminal embedding.
\end{lemma}

\begin{proof}
    Set $R(x) = d(x, V)$ and
    \[
        \hat\varphi(x) = \frac{1}{D} \tilde\varphi(x) \oplus_2 R(x).
    \]
    Since both $\tilde\varphi / D$ and $R$ are $1$-Lipschitz, $\Lip(\hat\varphi) \leq \sqrt{2}$.
    For $u \in V$ and $y \in X$, choose $v \in V$ closest to~$y$.
    Then
    \begin{align*}
        d(u, y)
         & \leq d(u, v) + R(y) \\
         & \leq \norm{\tilde\varphi(u) - \tilde\varphi(v)} + R(y) &  & \because\ \varphi \text{ is non-contracting on } V \\
         & \leq \norm{\tilde\varphi(u) - \tilde\varphi(y)} + \norm{\tilde\varphi(y) - \tilde\varphi(v)} + R(y) \\
         & \leq \norm{\tilde\varphi(u) - \tilde\varphi(y)} + (D + 1) R(y) &  & \because\ \Lip(\tilde\varphi) \leq D \\
         & \leq 3 D \, \norm{\hat\varphi(u) - \hat\varphi(y)} &  & \because\ R(u) = 0.
    \end{align*}
    Therefore, $\hat\varphi / \sqrt{2}$ is $1$-Lipschitz and contracts distances from $V$ by at most a factor of $3 \sqrt{2} D$, as required.
\end{proof}

\begin{lemma}[Lipschitz extension constant upper bounds distortion]\label{lem:extendability-upper}
    Let $(X, d)$ be a finite metric space covered by nonempty sets $U_1, \ldots, U_k$, so that each $U_i$ is Euclidean in $X$.
    Then $c_2(X) \leq O(\lipext_{\max} \log(k + 1))$.
\end{lemma}

\begin{proof}
    Fix $i \in [k]$ and an isometric embedding $\varphi_i : U_i \to \ell_2$.
    It is $1$-Lipschitz, so it extends to $\tilde\varphi_i : X \to \ell_2$ with $\Lip(\tilde\varphi_i) \leq \lipext(U_i) \leq \lipext_{\max}$, and \cref{lem:terminal-from-extension} gives a $3 \sqrt{2} \, \lipext_{\max}$-terminal embedding for $U_i$.
    Now \cref{thm:composition} gives the claim.
\end{proof}

\subsection{Lower Bound}

\begin{lemma}[Lipschitz extension constant lower bounds distortion]\label{lem:extendability-lower}
    Let $(X, d)$ be a metric space that embeds into $\ell_2$ with distortion $D$, and let $V \subseteq X$ be a nonempty subset.
    Then every Lipschitz map $f : V \to \ell_2$ extends to a map $\tilde f : X \to \ell_2$ with $\Lip(\tilde f) \leq D \Lip(f)$.
    In particular, $\lipext(V) \leq c_2(X)$.
\end{lemma}
\begin{proof}
    Let $F : X \to \ell_2$ be an embedding with distortion $D$, so that $d(x, y) \leq \norm{F(x) - F(y)} \leq D \, d(x, y)$ for all $x, y \in X$.
    Since $F$ does not contract distances, it is injective, and the map $g = f \circ F^{-1}$ is well-defined on $F(V) \subseteq \ell_2$.
    For $u, v \in V$,
    \[
        \norm{g(F(u)) - g(F(v))} = \norm{f(u) - f(v)} \leq \Lip(f) \, d(u, v) \leq \Lip(f) \, \norm{F(u) - F(v)},
    \]
    so $g$ is $\Lip(f)$-Lipschitz on $F(V)$.
    By Kirszbraun's Theorem~\cite{kirszbraun1934zusammenziehende}, the map $g$ extends to a map $G : \ell_2 \to \ell_2$ with $\Lip(G) \leq \Lip(f)$.
    Let $\tilde f = G \circ F$.
    We show that this is a Lipschitz extension of $f$.
    For $u \in V$ we have $\tilde f(u) = G(F(u)) = g(F(u)) = f(u)$, so, indeed, $\tilde f$ extends $f$.
    For $x, y \in X$,
    \[
        \norm{\tilde f(x) - \tilde f(y)} \leq \Lip(f) \, \norm{F(x) - F(y)} \leq D \Lip(f) \, d(x, y),
    \]
    which gives $\Lip(\tilde f) \leq D \Lip(f)$.
    The map $f$ was an arbitrary Lipschitz map for $V$, so $D$ must be at least $\lipext(V)$.
    Taking $D = c_2(X)$ gives $\lipext(V) \leq c_2(X)$.
\end{proof}

\begin{proof}[Proof of \cref{thm:extendability}]
    Discard empty sets among $\{U_i\}_{i = 1}^k$.
    The lower bound is \cref{lem:extendability-lower} applied to each $U_i$, and the upper bound is \cref{lem:extendability-upper}.
\end{proof}

\section{Lipschitz Extension Constants of Euclidean and 1-Strongly Euclidean Sets}
\label{sec:euclidean-unions}

In this section we prove \cref{lemma:lipschitz-constants}.
Throughout, $X$ is a finite pseudometric space, so distinct points may be at distance zero, and $U_1, \ldots, U_k$ are nonempty sets covering $X$ that are Euclidean, or $1$-strongly Euclidean, in $X$.
We fix $A = U_i$ and a $1$-Lipschitz map $f : A \to \ell_2$, and we construct an extension $F : X \to \ell_2$ of $f$ with $\Lip(F) = O(k^{\eta})$, where $\eta = \eta_0 = 1.1275\ldots$ for Euclidean sets and $\eta = \eta_1 = 0.8941\ldots$ for $1$-strongly Euclidean sets.

We bound $\Lip(F)$ through a metric extension problem.
Given a weighted graph $G$ on $X$, removing the vertices outside $A$ gives a graph $G^A$ on $A$ whose energy on $f$ is the least energy of an extension of $f$ (\cref{lem:canonical}), and the best quality it guarantees as a metric sparsifier, over all graphs $G$, is the square of the Lipschitz extension constant taken over all pseudometrics that are Euclidean on every $U_j$ (\cref{thm:euclidean-equivalence}).
We bound the quality by adding the sets $U_j$ back one at a time (\cref{lem:eliminate,lem:improved-one-point-energy}), turn the energy bounds into extensions (\cref{lem:improved-asymmetric-extension}), and average over the order in which the sets are added (\cref{lem:improved-averaging}).

\subsection{Euclidean metric sparsifiers and Lipschitz extensions}
\label{sec:euclidean-sparsifiers}

We obtain the Lipschitz extension bounds needed above by studying metric extension costs and their sparsifiers.
All sets in this section are finite, and we allow pseudometrics: distinct points may have distance zero.

In \emph{metric extension problems}, we are given a weighted graph $G$ on $X$ with nonnegative weights $g_{xy}=g_{yx}$ and $g_{xx}=0$, a nonempty set of terminals $A\subseteq X$, and a metric $\rho$ on $A$.
We seek a metric $d$ on $X$ extending $\rho$ that minimizes
\[
    \sum_{x<y}g_{xy}d(x,y)^p
\]
for a fixed exponent $p\ge1$, subject to problem-specific constraints
on $d$.
Here $x<y$ means that each unordered pair is included once.

For example, in the \emph{$0$-extension problem}, studied in
\cite{Kar98} and \cite{CKR01}, we take $p=1$ and require
\[
    d(x,y)=\rho(r(x),r(y)),
\]
where $r:X\to A$ fixes every terminal.
Equivalently, every vertex must be at distance zero from some terminal.
The standard LP relaxation drops this requirement and allows any pseudometric extension of $\rho$, with no further constraints.

In the \emph{metric labeling problem} of \cite{KT02}, we map the vertices into a prescribed ambient metric space $(Y,d_Y)$ and take $d(x,y)=d_Y(r(x),r(y))$, where $r:X\to Y$.
In the extension setting, $(Y,d_Y)$ contains $(A,\rho)$ as a subspace and $r$ fixes $A$, while the remaining vertices may be mapped anywhere in $Y$.
With $p=1$, the objective adds a nonnegative assignment cost $c(x,r(x))$ for each nonterminal vertex to the pairwise distance cost: $\sum_{x<y}g_{xy}d(x,y)+\sum_{x\in X\setminus A}c(x,r(x))$.

For a problem-specific class $\calD$ of metrics on $X$, define
the optimal extension cost by
\[
    \minext_{\calD}(\rho,G)
    =\inf_{\substack{d\in\calD\\d|_{A\times A}=\rho}}
    \sum_{x<y}g_{xy}d(x,y)^p.
\]
A metric sparsifier lets us approximate this cost using only the
distances on $A$, without computing an extension to $X$. Following
the formulation from \cite{MM16} for $p=1$, we call a weighted graph
$H$ on $A$, with nonnegative weights $h_{ab}$, a \emph{metric sparsifier}
of quality $Q\ge1$ if
\[
    \frac1Q\sum_{a<b}h_{ab}\rho(a,b)^p
    \le\minext_{\calD}(\rho,G)
    \le\sum_{a<b}h_{ab}\rho(a,b)^p
\]
for every terminal metric $\rho$ admitting an extension in $\calD$.
The same graph $H$ provides this approximation for every such $\rho$.

For our purposes, we fix a partition of $X$ into nonempty sets $U_1,\ldots,U_k$, take $A=U_i$ and $p=2$, and let $\calD$ be the class of pseudometrics $d$ on $X$ that are Euclidean on every $U_j$.
For any pseudometric $d$ on $X$, define its \emph{energy} by
\[
    \calE_G(d^2)=\sum_{x<y}g_{xy}d(x,y)^2.
\]
This definition does not require a Euclidean representation of $d$.
For a map $f:X\to\ell_2$, we also define
\[
    \calE_G(f)=\sum_{x<y}g_{xy}\|f(x)-f(y)\|^2.
\]
This is the energy of the pseudometric induced by $f$.
Thus, $\calE_G(f)=\calE_G(d^2)$ when $d(x,y)=\|f(x)-f(y)\|$ on all of $X$.
A general $d\in\calD$ need not admit such a map: only its restrictions to the individual sets $U_j$ are required to be Euclidean.
For real-valued $f$, $\calE_G(f)$ is the energy dissipated per unit time by the electrical network with conductances $g_{xy}$ and vertex potentials $f(x)$.
For $\ell_2$-valued $f$, it is the sum of the coordinate energies.

Specializing the definitions above, for a Euclidean metric $\rho$ on
$A$ let
\begin{equation}\label{eq:minext}
    \minext_{\calD}(\rho,G)
    =\inf_{\substack{d\in\calD\\d|_{A\times A}=\rho}}\calE_G(d^2).
\end{equation}

\begin{definition}\label{def:sparsifier}
    A weighted graph $H$ on $A$, with nonnegative weights, is a
    \emph{Euclidean metric sparsifier of $G$ for $\calD$ with quality $Q\ge1$}
    if, for every Euclidean $\rho$ on $A$,
    \begin{equation}\label{eq:sparsifier}
        \frac{1}{Q}\calE_H(\rho^2)
        \le\minext_{\calD}(\rho,G)\le\calE_H(\rho^2).
    \end{equation}
\end{definition}

The following theorem makes the connection with Lipschitz extension constants precise.

Let $\mathsf Q(\calD,A)$ be the least quality guaranteed for every weighted graph $G$ on $X$.
Let $\lipext(\calD,A)$ be the least $\Lambda\ge1$ such that, for every $d\in\calD$, every Lipschitz map $f:(A,d)\to\ell_2$ extends to $F : (X,d)\to\ell_2$ with $\Lip(F)\le\Lambda\Lip(f)$.
The graph and the metric may vary, while the class $\calD$ and the distinguished set $A$ are fixed.
Let $\mathsf Q_k$ and $\lipext_k$ be the suprema of these constants over all finite partitions into at most $k$ sets and all choices of $A$.
For the pseudometric $d$ of $X$ itself, which belongs to $\calD$, the constant $\lipext(A)$ of \cref{sec:introduction} is the value at $d$ of the supremum defining $\lipext(\calD,A)$, so
\begin{equation}
    \lipext(A)\le\lipext(\calD,A)\le\lipext_k.
\end{equation}
Hence, bounding $\lipext_k$ allows us to bound $\lipext(A)$.
We have the following result connecting the quality $\mathsf Q_k$ and the supremum $\lipext_k$.
The proof is deferred to \cref{app:equivalence}.

\begin{theorem}\label{thm:euclidean-equivalence}
    For every partition $U_1, \ldots, U_k$ of $X$, with $\calD$ as above, and every $A = U_i$, we have $\mathsf Q(\calD, A)=\lipext(\calD, A)^2$.
    As a consequence, $\mathsf Q_k = \lipext_k^2$.
\end{theorem}

\begin{remark}
    The same connection between sparsifier quality and Lipschitz
    extension constants holds for other metric extension problems.
    For example, \cite[Theorems 5.1--5.2]{MM16} showed that the optimal qualities of cut and flow sparsifiers equal the corresponding Lipschitz extension constants between suitable Banach spaces.
    We omit the details here.
\end{remark}

\subsection{The canonical Euclidean metric sparsifier}

We first consider the extension problem in which all of $X$ must be Euclidean.
Fix a nonempty set $A\subseteq X$.
We define $G^A$ to be
the graph on $A$ with nonnegative weights whose energy gives the
minimum Euclidean extension cost:
\[
    \calE_{G^A}(\rho^2)
    =\min_{\substack{d|_{A\times A}=\rho\\
            d\text{ Euclidean on }X}}\calE_G(d^2)
\]
for every Euclidean pseudometric $\rho$ on $A$.
Only the distances on $A$ are fixed.
All other distances can be chosen to minimize the cost.
We will show that one graph $G^A$ works for every $\rho$, and that its weights are uniquely determined.
The sets $U_j$ play no role here.
We return to them when comparing this minimum with the cost of extensions in $\calD$.

The weights of $G^A$ can be found by removing one vertex at a time.
To see how, suppose we first remove a nonisolated vertex $v$ with neighbors $x_1,\ldots,x_t$.
Fix a Euclidean pseudometric $d$ on $X\setminus\{v\}$.
The distances among the neighbors can be written as $d(x_a,x_b)=\|u_a-u_b\|$ for suitable vectors $u_1,\ldots,u_t$.
Choosing a position $z$ for $v$ adds $\sum_{r=1}^t g_{vx_r}\|z-u_r\|^2$ to the energy.
Recall that this is the objective for a single cluster in the weighted $k$-means problem.
The optimal choice of the center is simply the
weighted mean
\[
    \bar u=\frac{1}{\delta_v}\sum_{r=1}^t g_{vx_r}u_r,
    \text{ where }
    \delta_v=\sum_{r=1}^t g_{vx_r}>0.
\]
Using the standard alternative formula for the cost of a cluster under
the $k$-means objective, we express the contribution of $v$ to the energy
without referring to $z$ explicitly as
\[
    \frac{1}{\delta_v}\sum_{1\le r<s\le t}
    g_{vx_r}g_{vx_s}\|u_r-u_s\|^2
    =\frac{1}{\delta_v}\sum_{1\le r<s\le t}
    g_{vx_r}g_{vx_s}d(x_r,x_s)^2.
\]
To capture the contribution of $v$ in $G'=G^{X\setminus\{v\}}$, we add
$g_{xv}g_{yv}/\delta_v$ to each weight $g_{xy}$:
\[
    G'_{xy}=g_{xy}+\frac{g_{xv}g_{yv}}{\delta_v},
\]
where $x,y\in X\setminus\{v\},\ x\ne y$.
The energy of $G'$ therefore includes the cost of the best position for $v$, even though $v$ is no longer in the graph.
We repeat this step until only $A$ remains, using the current weights and the current value of $\delta_v$ at each step.
An isolated vertex is simply deleted.
This gives the graph $G^A$ described above, as we now prove.

\begin{lemma}\label{lem:canonical}
    For every weighted graph $G$ on $X$ and nonempty $A\subseteq X$, the
    graph $G^A$ defined above exists and is unique.
    The update rule above constructs it by removing the vertices of $X\setminus A$ in any order.
    For every map $f:A\to\ell_2$, it satisfies
    \begin{equation}\label{eq:minimum-energy}
        \calE_{G^A}(f)
        =\min_{\substack{F:X\to\ell_2\\
                F|_A=f}}\calE_G(F).
    \end{equation}
    Moreover, for $A \subseteq A' \subseteq X$ we have:
    \begin{equation}\label{eq:nested}
        (G^{A'})^A = G^A.
    \end{equation}
    For $A=U_i$, allowing all extensions in $\calD$ gives $\minext_{\calD}(\rho,G)\le\calE_{G^A}(\rho^2)$ for every Euclidean $\rho$ on $A$, which is the right inequality in \eqref{eq:sparsifier}.
\end{lemma}

\begin{proof}
    We begin by turning the weighted $k$-means discussion into a proof
    that the update rule gives the graph we want.
    Fix positions $f(a)\in\ell_2$ for the vertices $a\in A$, and remove the other vertices one at a time.
    With the positions of the remaining vertices fixed, the calculation above replaces the cost of the best position for $v$ by the added edge weights.
    Thus, after each removal, the current graph records the minimum energy over the positions of all vertices removed so far.
    When only $A$ remains, the resulting graph
    $H$ satisfies
    \[
        \calE_H(f)=\min_{\substack{F:X\to\ell_2\\
                F|_A=f}}\calE_G(F).
    \]
    We can obtain a minimizing extension by restoring the removed vertices
    in reverse order, placing each at the weighted mean of its neighbors
    with the weights used when it was removed.
    An isolated vertex can be placed anywhere, since its position does not affect the energy.

    Now let $\rho$ be a Euclidean pseudometric on $A$, and choose $f$ with $\rho(a,b)=\|f(a)-f(b)\|$.
    Every Euclidean extension of $\rho$ can be represented by vectors that agree with $f$ on $A$, so the minimum above is exactly the minimum Euclidean extension cost of $\rho$.
    Thus, $H$ has the defining property of $G^A$, and \eqref{eq:minimum-energy} holds.
    Its weights are nonnegative, and the same graph works for every $\rho$, since the updates depend only on the graph weights.

    These energies also determine the graph uniquely.
    Indeed, for real-valued $f$, the coefficient of $f(x)f(y)$ in $\calE_H(f)$ is $-2h_{xy}$ for $x\ne y$.
    Hence, every removal order gives the same graph $G^A$.
    In particular, we may first remove all vertices outside $A'$ and then those in $A'\setminus A$.
    This proves $(G^{A'})^A=G^A$.

    Finally, when $A=U_i$, every globally Euclidean extension of $\rho$ belongs to $\calD$.
    Allowing all extensions in $\calD$ can only lower the minimum cost, so $\minext_{\calD}(\rho,G)\le\calE_{G^A}(\rho^2)$, as claimed.
\end{proof}

It remains to compare $\calE_{G^A}(\rho^2)$ with the energy of extensions of $\rho$ that are Euclidean only on the individual sets $U_j$.
We eliminate the sets $U_j$ with $j\ne i$ one at a time, and first bound the energy added when a single Euclidean set is eliminated.
We use $G[B]$ to denote the graph induced by $B$, with the original weights.

\subsection{A matrix estimate}

The next claim helps bound the cost of the new edges created when we remove a Euclidean set $B$.
It controls weighted averages of squared distances within $B$ using the energy on its internal edges.
In the application, the vertex weight $s_b$ will be the total weight of edges from $b\in B$ to the vertices we keep.
We let $L$ denote the non-normalized Laplacian of a weighted graph with edge weights $w_{bb'}$.
For vectors $z_b$ in a Hilbert space, one for
each vertex, we have
\[
    \sum_b\ip{z_b}{(Lz)_b}
    =\sum_{b<b'}w_{bb'}\norm{z_b-z_{b'}}^2.
\]

\begin{claim}\label{clm:averaging}
    Let $L$ be the Laplacian of a weighted graph on $B$.
    Assign a positive weight $s_b$ to each vertex $b\in B$, and let $C=\diag(s_b)$, $K=C^{-1}L$, and $P(t)=e^{-tK}\quad(t\ge0)$.
    Each $P(t)$ has nonnegative entries and satisfies $P(t)\one=\one$, $CP(t)=P(t)^{\mathsf T}C$ and $P(t)^{\mathsf T}s=s$, where $\one$ is the all-ones vector and $s=(s_b)_{b\in B}$ is the vector of vertex weights.
    The first two identities say that the rows of $P(t)$ sum to one and that $CP(t)$ is symmetric.
    For vectors $z_b$ in a Hilbert space,
    \begin{equation}\label{eq:averaged-distances}
        \sum_{b,b'}s_bP_{bb'}(t)\norm{z_b-z_{b'}}^2
        \le 2t\sum_{b<b'}w_{bb'}\norm{z_b-z_{b'}}^2.
    \end{equation}
\end{claim}

\begin{proof}
    The nonnegative entries and the identity $P(t)\one=\one$ follow by
    writing $P(t)$ as the limit of $(\Id-tK/N)^N$: for sufficiently large
    $N$, the matrix $\Id-tK/N$ has nonnegative entries and fixes $\one$.

    With respect to the inner product
    \[
        \ip{u}{v}_C=\sum_b s_b\ip{u_b}{v_b},
    \]
    the matrix $K$ is self-adjoint and nonnegative, since $CK=L$.
    Thus, $P(t)$ is also self-adjoint, so $CP(t)=P(t)^{\mathsf T}C$.
    Since $s=C\one$, we also get
    \[
        P(t)^{\mathsf T}s=P(t)^{\mathsf T}C\one
        =CP(t)\one=C\one=s.
    \]

    Expanding the squared distances now gives
    \begin{align*}
        \sum_{b,b'}s_bP_{bb'}(t)\norm{z_b-z_{b'}}^2
         & =2\ip{z}{(\Id-P(t))z}_C \\
         & \le 2t\ip{z}{Kz}_C
        =2t\sum_{b<b'}w_{bb'}\norm{z_b-z_{b'}}^2.
    \end{align*}
    The inequality follows by diagonalizing $K$ and applying
    $1-e^{-t\lambda}\le t\lambda$ to each eigenvalue $\lambda\ge0$.
\end{proof}

\subsection{Eliminating one Euclidean set}

\begin{lemma}\label{lem:eliminate}
    Let $Y=A\sqcup B$ be a finite pseudometric space, where $A\ne\varnothing$ and
    $B$ is Euclidean.
    For a weighted graph $G$ on $Y$, let $I$ be the energy on edges within $B$ and $J$ the energy on edges between $A$ and $B$, so $I = \calE_{G[B]} (d^2)$ and $J = \sum_{a \in A, b\in B} g_{ab} d(a, b)^2$.
    Then
    \begin{equation}\label{eq:eliminate}
        \calE_{G^A}(d^2)
        \le\calE_G(d^2)+J+\sqrt{2\pi IJ}.
    \end{equation}
\end{lemma}

\begin{proof}
    The idea is to bound the distance across each new edge on $A$ by three distances: from $A$ into $B$, across $B$, and back to $A$.
    The two outside parts will be controlled by $J$, and \cref{clm:averaging} will control the middle part using $I$.

    For $b\in B$, take $s_b=\sum_{a\in A}g_{ab}$.
    We first assume that $B$ is nonempty and $s_b>0$ for every $b\in B$, so each vertex of $B$ has an edge to $A$.
    We remove this assumption at the end.
    Since $B$ is Euclidean, choose vectors $z_b$ with $d(b,b')=\norm{z_b-z_{b'}}$.

    We need to account for the weights added on $A$ when $B$ is removed.
    To compute them, we give each edge from $A$ to $B$ its own copy of its endpoint in $A$.
    Each copy then has just one neighbor in $B$, which makes the new weights easy to compute.
    More precisely, for each positive-weight edge with endpoints $a_p\in A$ and $b_p\in B$, create a vertex $p$ and join it to $b_p$ with the original edge weight $c_p$.
    Our auxiliary graph consists of $G[B]$ and these new edges.
    Remove $B$ from this graph, and call the resulting weights between the copies $\omega_{pq}$.

    To recover the weights added on $A$, give all copies of each vertex the same real value and use \eqref{eq:minimum-energy}.
    The weight added between two vertices of $A$ is the sum of $\omega_{pq}$ over their copies.
    Consequently,
    \begin{equation}\label{eq:added-cost}
        \calE_{G^A}(d^2)-\calE_{G[A]}(d^2)
        =\sum_{p<q}\omega_{pq}d(a_p,a_q)^2.
    \end{equation}
    This is an identity between edge weights, so it holds even though
    $A$ need not be Euclidean.

    Let $L$ be the Laplacian of $G[B]$, and let $C=\diag(s_b)$ and $M = L + C$.
    Since all $s_b$ are positive, $M$ is positive definite.
    Minimizing
    over the values at $B$ in the auxiliary graph gives for $p \not= q$
    \begin{equation}\label{eq:new-weights}
        \omega_{pq}=c_pc_q(M^{-1})_{b_pb_q}.
    \end{equation}
    Indeed, if the copies have real values $\eta_p$ and the vertices of $B$
    have real values $\xi_b$, the energy is
    \[
        \xi^{\mathsf T}M\xi-2\sum_p c_p\eta_p\xi_{b_p}+\sum_p c_p\eta_p^2.
    \]
    Completing the square and comparing the coefficients of $\eta_p\eta_q$
    gives \eqref{eq:new-weights}.

    We next express these weights using the matrices $P(t)$ from \cref{clm:averaging}, applied to $G[B]$ with vertex weights $s_b$.
    With $K=C^{-1}L$ and $P(t)=e^{-tK}$, we have
    \[
        M^{-1}=(\Id+K)^{-1}C^{-1}
        =\int_0^\infty e^{-t}P(t)C^{-1}\dd t.
    \]
    Thus, \eqref{eq:new-weights} gives
    $\omega_{pq}=\int_0^\infty e^{-t}\mu_{pq}(t)\dd t$ for $p\ne q$, where
    \[
        \mu_{pq}(t)=c_pP_{b_pb_q}(t)\frac{c_q}{s_{b_q}}.
    \]
    These weights are nonnegative and symmetric, and satisfy
    \begin{equation}\label{eq:pair-weights}
        \sum_q\mu_{pq}(t)=c_p,\qquad \sum_p\mu_{pq}(t)=c_q.
    \end{equation}
    For the first equality, group the copies $q$ by their neighbor $b_q$,
    use $\sum_{q:b_q=b'}c_q=s_{b'}$, and then use the row sums of $P(t)$.
    The second follows by symmetry, which comes from $s_bP_{bb'}(t)=s_{b'}P_{b'b}(t)$.

    For each $t\ge0$, let
    \[
        S_t=\sum_{p,q}\mu_{pq}(t)d(a_p,a_q)^2.
    \]
    By symmetry, we may replace the sum over $p<q$ in
    \eqref{eq:added-cost} by half the sum over all $p,q$.
    The terms with $p=q$ contribute zero.
    Hence,
    \begin{equation}\label{eq:added-cost-integral}
        \calE_{G^A}(d^2)-\calE_{G[A]}(d^2)
        =\frac12\int_0^\infty e^{-t}S_t\dd t.
    \end{equation}

    We can now use the three distances along the route $a_p\to b_p\to b_q\to a_q$.
    Let $u_p=d(a_p,b_p)$.
    The triangle
    inequality gives
    \[
        d(a_p,a_q)\le u_p+\norm{z_{b_p}-z_{b_q}}+u_q.
    \]
    By \eqref{eq:pair-weights}, the two outside distances satisfy
    \[
        \sum_{p,q}\mu_{pq}(t)u_p^2
        =\sum_{p,q}\mu_{pq}(t)u_q^2
        =\sum_p c_pu_p^2=J.
    \]
    For the middle distance, group the copies by their neighbors in $B$
    and apply \eqref{eq:averaged-distances}:
    \[
        \sum_{p,q}\mu_{pq}(t)\norm{z_{b_p}-z_{b_q}}^2
        =\sum_{b,b'}s_bP_{bb'}(t)\norm{z_b-z_{b'}}^2
        \le 2tI.
    \]
    The triangle inequality for the Euclidean norm with weights
    $\mu_{pq}(t)$ therefore gives
    \[
        \sqrt{S_t}\le\sqrt J+\sqrt{2tI}+\sqrt J.
    \]
    Substituting into \eqref{eq:added-cost-integral}, we obtain
    \begin{align*}
        \calE_{G^A}(d^2)-\calE_{G[A]}(d^2)
         & \le\frac12\int_0^\infty e^{-t}
        \bigl(4J+4\sqrt{2tIJ}+2tI\bigr)\dd t \\
         & =2J+\sqrt{2\pi IJ}+I,
    \end{align*}
    where we used $\int_0^\infty e^{-t}\sqrt t\dd t=\sqrt\pi/2$.
    Since $\calE_G(d^2)=\calE_{G[A]}(d^2)+I+J$, this proves \eqref{eq:eliminate} under our assumption that all $s_b$ are positive.

    To remove that assumption, let $B_0=\{b\in B:s_b>0\}$.
    If $B_0=\varnothing$, no edges join $A$ to $B$, so $G^A=G[A]$ and the bound is immediate.
    Otherwise, first remove the vertices of $B\setminus B_0$.
    Throughout this process their neighbors lie in $B$, so the edges within $A$ and the edges from $A$ to $B_0$ keep their weights.
    Let $G'$ be the resulting graph on $A\sqcup B_0$.
    By \eqref{eq:nested}, $(G')^A=G^A$, and the graph on $B_0$ is $(G[B])^{B_0}$.
    Applying \eqref{eq:minimum-energy} to vectors representing the distances on $B$, we find that the internal and crossing energies of $G'$ satisfy $I'=\calE_{G'[B_0]}(d^2)\le I$ and $J' = J$.
    Thus, $\calE_{G'}(d^2)\le\calE_G(d^2)$, and every vertex of $B_0$ still has an edge to $A$.
    The bound already proved for $G'$ therefore gives \eqref{eq:eliminate} for $G$ as well.
\end{proof}

\subsection{From energy bounds to Lipschitz extensions}
\label{sec:improved-extension-bounds}

We now turn the one-set estimate into a bound on the Lipschitz constant of an extension.
Two factors matter separately: the growth on pairs of points inside the eliminated set, and the growth on pairs crossing it.
We first show that an energy bound with separate coefficients for the internal and the crossing energy yields an extension with separate Lipschitz factors on the two kinds of pairs (\cref{lem:improved-asymmetric-extension}).
Averaging these extensions over the choice of the set added last then reduces the growth at the step with $r$ sets to a factor of $1 + \eta / r$ (\cref{lem:improved-averaging}).
Finally, we choose the two factors to minimize the exponent $\eta$.

We write $H_0$ for the assumption that every $U_i$ is Euclidean in $X$, and $H_1$ for the assumption that every $U_i$ is $1$-strongly Euclidean in $X$, that is, $U_i \cup \{x\}$ is Euclidean for every $x \in X$.
Both conditions are preserved when a set or the ambient space is restricted.
Under $H_1$, we also sharpen the one-set estimate \eqref{eq:eliminate}.

\begin{lemma}\label{lem:improved-asymmetric-extension}
    Let $Y = A \sqcup B$ be a finite pseudometric space with $A \ne \varnothing$, and let $t,s \ge 1$. Suppose that every nonnegatively weighted graph $G$ on $Y$ with no edges inside $A$ satisfies
    \[
        \calE_{G^A}(d^2) \le t^2 I + s^2 J,
    \]
    where $I$ and $J$ are as in \cref{lem:eliminate}. Then every $1$-Lipschitz map $f:A \to \ell_2$ has an extension $F:Y \to \ell_2$ such that
    \[
        \norm{F(b)-F(b')} \le t d(b,b'), \qquad \norm{F(a)-F(b)} \le s d(a,b),
    \]
    where $a \in A,\ b,b' \in B$.
    The extension can be chosen with values in $\conv(f(A))$.
\end{lemma}

\begin{proof}
    The case $B = \varnothing$ is immediate. Otherwise, by \cref{lem:canonical} and the nonnegative weights of $G^A$,
    \[
        \min_{F|_A = f} \calE_G(F) = \calE_{G^A}(f) \le \calE_{G^A}(d^2) \le t^2 I + s^2 J.
    \]
    Projection onto the compact convex set $\conv(f(A))$ fixes $f(A)$ and does not increase distances, so this inequality remains valid when all extension values are restricted to that set.
    Let $\mathcal F$ be the resulting compact convex set of extensions, and let $\mathcal W$ be the probability simplex on unordered pairs with at least one endpoint in $B$.
    For such a pair, let $b_{xy} = t^2$ if $x,y \in B$ and $b_{xy} = s^2$ otherwise, and define
    \[
        \Phi(F,w) = \sum_{\{x,y\}} w_{xy} \bigl(\norm{F(x)-F(y)}^2 - b_{xy} d(x,y)^2\bigr).
    \]
    The energy inequality gives $\min_{F \in \mathcal F} \Phi(F,w) \le 0$ for every $w \in \mathcal W$. As in \cref{app:equivalence}, convexity in $F$ and affinity in $w$ give
    \[
        \min_{F \in \mathcal F} \max_{w \in \mathcal W} \Phi(F,w) = \max_{w \in \mathcal W} \min_{F \in \mathcal F} \Phi(F,w) \le 0.
    \]
    Taking $w$ supported on each individual pair proves the two bounds, including pairs at pseudodistance zero.
\end{proof}

We now average such extensions over the choice of the set added last.

\begin{lemma}\label{lem:improved-averaging}
    Suppose that an added set satisfying a given geometric condition always admits the extension in \cref{lem:improved-asymmetric-extension} with fixed factors $t,s \ge 1$, and that the condition is preserved by restriction.
    Let $\eta = \max\{t-1,2(s-1)\}$.
    If $A \ne \varnothing$ and $X \setminus A$ is partitioned into $m$ such sets, every $1$-Lipschitz map $f:A \to \ell_2$ extends to $X$ with Lipschitz constant at most
    \begin{equation}\label{eq:improved-extension-product}
        P_m = \prod_{r = 1}^m \left(1+\frac{\eta}{r}\right).
    \end{equation}
    For fixed $\eta$, this is $O((m+1)^\eta)$.
    If $\eta \ge 1$, it is at most $(m+1)^\eta$.
\end{lemma}

\begin{proof}
    We induct on $m$, starting with $P_0 = 1$.
    Let $X \setminus A = B_1 \sqcup \cdots \sqcup B_m$.
    For each $i \in [m]$, first extend the prescribed map to $X \setminus B_i$ with Lipschitz constant at most $P_{m-1}$, and then add $B_i$ using \cref{lem:improved-asymmetric-extension}, after scaling by $P_{m-1}$.
    Call the resulting map $F_i$.
    All these maps take values in the same Hilbert space and agree with the prescribed map on $A$, so their arithmetic average
    \[
        F(x) = \frac{1}{m} \sum_{i = 1}^m F_i(x)
    \]
    also extends that map.
    For a pair inside one $B_j$, only $F_j$ uses the factor $t$.
    For a pair in two distinct complementary sets, exactly two summands use the factor $s$.
    A pair joining $A$ to $B_j$ has only one summand using $s$, and pairs inside $A$ are unchanged.
    The triangle inequality therefore gives
    \[
        \Lip(F) \le P_{m-1} \max\left\{1+\frac{t-1}{m},\ 1+\frac{2(s-1)}{m}\right\} = P_m.
    \]
    Finally, for $m \ge 1$,
    \[
        P_m \le \exp\left(\eta \sum_{r = 1}^m \frac{1}{r}\right) \le e^\eta (m+1)^\eta.
    \]
    When $\eta \ge 1$, the sharper estimate follows by multiplying $1+\eta/r \le (1+1/r)^\eta$.
    For $0 < \eta < 1$, we use the preceding exponential estimate instead.
\end{proof}

\subsection{The exponent for Euclidean sets}

For a graph with no edges inside $A$, the estimate \eqref{eq:eliminate} is
\[
    \calE_{G^A}(d^2) \le I + 2J + \sqrt{2\pi IJ}.
\]
Consequently, \cref{lem:improved-asymmetric-extension} applies whenever
\begin{equation}\label{eq:improved-h0-tradeoff}
    t \ge 1, \qquad s \ge \sqrt{2}, \qquad (t^2-1)(s^2-2) \ge \frac{\pi}{2},
\end{equation}
since $\sqrt{2\pi IJ} \le (t^2-1)I + (s^2-2)J$ under this condition.
To minimize the exponent supplied by \cref{lem:improved-averaging}, take $t = 1+\eta$ and $s = 1+\eta/2$: for a fixed permitted exponent $\eta$, these are the largest available factors.

Let $\eta_0 > 1$ be the unique root of
\begin{equation}\label{eq:improved-eta-zero}
    \eta_0(\eta_0+2)(\eta_0^2+4\eta_0-4) = 2\pi.
\end{equation}
The left side is strictly increasing for $\eta_0 \ge 1$, equals $3$ at $1$, and tends to infinity.
Numerically, $\eta_0 = 1.1275681325\ldots$.
The choices $t = 1+\eta_0$ and $s = 1+\eta_0/2$ give equality in \eqref{eq:improved-h0-tradeoff}.
Thus, a complement consisting of $m$ Euclidean sets admits extension factor at most $(m+1)^{\eta_0}$.
The simpler exponent $8/7$ is also valid, since $t = 15/7$ and $s = 11/7$ give $(t^2-1)(s^2-2) = 4048/2401 > \pi/2$.

\subsection{Eliminating one $1$-strongly Euclidean set}

\begin{lemma}\label{lem:improved-one-point-energy}
    In the setting of \cref{lem:eliminate}, suppose that $B \cup \{a\}$ is Euclidean for every $a \in A$. Then
    \begin{equation}\label{eq:improved-one-point-energy}
        \calE_{G^A}(d^2) \le \calE_{G[A]}(d^2) + \frac{1}{2}\left(\sqrt{J}+\sqrt{J+I+\sqrt{I(I+J)}}\right)^2.
    \end{equation}
\end{lemma}

\begin{proof}
    Use the notation $s_b,c_p,a_p,b_p,C,K,P(t),\mu_{pq}(t)$ from the proof of \cref{lem:eliminate}, and first assume that $s_b > 0$ for every $b \in B$. Let
    \[
        R = (\Id+K)^{-1} = \int_0^\infty e^{-t}P(t)\dd t,
    \]
    and
    \[
        \nu_{pq} = \int_0^\infty e^{-t}\mu_{pq}(t)\dd t = c_p R_{b_p b_q}\frac{c_q}{s_{b_q}}.
    \]
    These weights are symmetric and have marginals $\sum_q \nu_{pq} = c_p$ and $\sum_p \nu_{pq} = c_q$. Equation~\eqref{eq:added-cost-integral} gives
    \[
        2\bigl(\calE_{G^A}(d^2)-\calE_{G[A]}(d^2)\bigr) = \sum_{p,q} \nu_{pq} d(a_p,a_q)^2.
    \]
    Choose a tight Euclidean realization $z_b$ of $B$. By the one-point condition and \cref{fact}, for each $a \in A$ there are a vector $u_a$ and a height $h_a \ge 0$ such that for every $b \in B$ we have
    \[
        d(a,b)^2 = \norm{u_a-z_b}^2+h_a^2.
    \]
    Set $m = \ip{z}{(\Id-R)z}_C$, $v = \ip{(\Id-R)z}{(\Id-R)z}_C$, and $T = \sum_{p,q} \nu_{pq} d(a_p,b_q)^2$. Expanding the squared distances and using the marginals gives
    \begin{align*}
        T-J & = 2\sum_p c_p \ip{u_{a_p}}{((\Id-R)z)_{b_p}} \\
         & = 2m + 2\sum_p c_p \ip{u_{a_p}-z_{b_p}}{((\Id-R)z)_{b_p}} \\
         & \le 2m + 2\sqrt{Jv},
    \end{align*}
    where the averaged squared norms of $z_b$ and the height terms cancel, and $\sum_p c_p \norm{u_{a_p}-z_{b_p}}^2 \le J$.

    Let $w_\lambda$ be the squared norms of the spectral components of $z$ for $K$ in the $C$ inner product.
    Then
    \[
        I = \sum_\lambda \lambda w_\lambda, \qquad m = \sum_\lambda \frac{\lambda}{1+\lambda}w_\lambda, \qquad v = \sum_\lambda \frac{\lambda^2}{(1+\lambda)^2}w_\lambda.
    \]
    Cauchy--Schwarz gives $m^2 \le I(m-v)$, and $0 \le m \le I$.
    If $I > 0$, then $m = I(1+q)/2$ with $-1 \le q \le 1$.
    Then
    \[
        2m+2\sqrt{Jv} \le I+Iq+\sqrt{IJ}\sqrt{1-q^2} \le I+\sqrt{I(I+J)}.
    \]
    If $I = 0$, the spectral expressions give $m = v = 0$, so in either case $T \le J+I+\sqrt{I(I+J)}$. Finally, the triangle inequality along $a_p,b_q,a_q$ and the Euclidean norm with weights $\nu_{pq}$ give
    \[
        \left(\sum_{p,q} \nu_{pq}d(a_p,a_q)^2\right)^{1/2} \le \sqrt{T}+\left(\sum_{p,q} \nu_{pq}d(a_q,b_q)^2\right)^{1/2} = \sqrt{T}+\sqrt{J}.
    \]
    This proves \eqref{eq:improved-one-point-energy}.
    If some $s_b$ vanish, use the reduction at the end of the proof of \cref{lem:eliminate}.
    It preserves $\calE_{G[A]}(d^2)$ and $J$, decreases $I$, and preserves the one-point condition, and the right side of \eqref{eq:improved-one-point-energy} is increasing in $I$.
    If all $s_b$ vanish, $G^A = G[A]$ and the conclusion is immediate.
\end{proof}

\subsection{The exponent for $1$-strongly Euclidean sets}

To use \cref{lem:improved-averaging} with exponent $\eta$, we again take $t = 1+\eta$ and $s = 1+\eta/2$.
For $I+J > 0$, set $z = \sqrt{1-\sqrt{I/(I+J)}}$.
Then
\[
    \frac{I}{I+J} = (1-z^2)^2, \qquad \frac{J}{I+J} = z^2(2-z^2), \qquad 0 \le z \le 1,
\]
and the added-energy bound in \eqref{eq:improved-one-point-energy}, divided by $I+J$, equals $\frac{1}{2}(z+1)^2(2-z^2)$. Thus, that bound is at most $t^2 I+s^2 J$ for all $I,J \ge 0$ precisely when
\begin{equation}\label{eq:improved-one-point-gap}
    D_\eta(z) = (1+\eta)^2(1-z^2)^2 + (1+\eta/2)^2 z^2(2-z^2) - \frac{1}{2}(z+1)^2(2-z^2) \ge 0 \qquad(0 \le z \le 1).
\end{equation}
Define the exponent exactly by
\begin{equation}\label{eq:improved-eta-one}
    \eta_1 = \min\{\eta \ge 0 : D_\eta(z) \ge 0 \text{ for every } z \in [0,1]\}.
\end{equation}
The minimum exists: $D_2(z) \ge 0$ on $[0,1]$, the condition is closed, and $D_\eta(z)$ is strictly increasing in $\eta \ge 0$ for every $z$.
This is the same optimization of the internal and crossing factors as under $H_0$.

For completeness, we certify the useful numerical interval
\[
    0.8941 < \eta_1 < 0.89417.
\]
Take $a = \frac{1}{2}+\eta+\frac{3}{4}\eta^2$. Expanding the gap gives
\begin{align*}
    D_\eta(z) & = \eta^2+2\eta-2z+\frac{z^2}{2}+z^3+a(z^4-2z^2) \\ D_\eta'(z) &= (z+1)\bigl(4az^2+(3-4a)z-2\bigr).
\end{align*}
Take $\eta = 89417/100000$ and $z_0 = 90189211/10^8$.
On $[0,3/4]$, the bounds $z^3 \le 9z/16$ and $z^2 \le 3z/4$ give $D_\eta'(z) \le -2+(13/4-7a/4)z < 0$.
On $[3/4,1]$,
\[
    D_\eta''(z) = 1+6z+12az^2-4a \ge \frac{11}{2}+\frac{11a}{4} > 10.
\]
Exact rational substitution gives $D_\eta(z_0) > 9 \cdot 10^{-8}$ and $|D_\eta'(z_0)| < 10^{-7}$. Consequently, throughout this latter interval,
\[
    D_\eta(z) \ge D_\eta(z_0)+D_\eta'(z_0)(z-z_0)+5(z-z_0)^2 \ge D_\eta(z_0)-\frac{D_\eta'(z_0)^2}{20} > 0.
\]
Monotonicity handles $[0,3/4]$, and strict positivity on the compact interval gives $\eta_1 < 0.89417$. For $\eta = 8941/10000$, direct substitution yields
\[
    2D_\eta(9/10) = -\frac{300526077}{2000000000000} < 0,
\]
so $\eta_1 > 0.8941$.

\subsection{Proof of Lemma~\ref{lemma:lipschitz-constants}}

\begin{proof}[Proof of \cref{lemma:lipschitz-constants}]
    Fix $i \in [k]$ and a Lipschitz map $f : U_i \to \ell_2$.
    A constant map extends constantly, and scaling handles other Lipschitz constants, so we may assume that $\Lip(f) = 1$.
    Keep $U_i$ whole and partition its complement into the sets $B_j = U_j \setminus (U_i \cup U_1 \cup \cdots \cup U_{j - 1})$ for $j \neq i$, discarding empty ones.
    This gives $m \leq k - 1$ sets, and each $B_j$ is Euclidean, respectively $1$-strongly Euclidean, in $X$, because both conditions are preserved by restriction.
    For Euclidean sets, \eqref{eq:eliminate} and \eqref{eq:improved-h0-tradeoff} show that \cref{lem:improved-asymmetric-extension} applies with $t = 1 + \eta_0$ and $s = 1 + \eta_0 / 2$.
    For $1$-strongly Euclidean sets, \cref{lem:improved-one-point-energy} and \eqref{eq:improved-eta-one} show that it applies with $t = 1 + \eta_1$ and $s = 1 + \eta_1 / 2$.
    In either case $\max\{t - 1, 2(s - 1)\} = \eta_j$, so \cref{lem:improved-averaging} extends $f$ to $X$ with Lipschitz constant at most $P_m \leq P_{k - 1}$, with $P_m$ as in \eqref{eq:improved-extension-product}.
    Hence,
    \[
        \lipext(U_i) \leq \prod_{r = 1}^{k - 1} \left(1 + \frac{\eta_j}{r}\right).
    \]
    Since $\eta_0 > 1$, this is at most $k^{\eta_0}$ under $H_0$, and since $\eta_1 < 1$, it is at most $e^{\eta_1} k^{\eta_1}$ under $H_1$.
\end{proof}

The proof applies to every $d \in \calD$ and every partition into at most $k$ sets, with the same product, so it also bounds the uniform constants of \cref{sec:euclidean-sparsifiers}.

\begin{corollary}\label{cor:uniform-constants}
    For every $k \geq 1$, we have $\lipext_k \leq \prod_{r = 1}^{k - 1} \left(1 + \frac{\eta_0}{r}\right) \leq k^{\eta_0}$, and $\mathsf Q_k = \lipext_k^2 \leq k^{2 \eta_0}$.
    In particular, for every partition of $X$ into at most $k$ Euclidean sets and every weighted graph $G$ on $X$, the canonical graph $G^{U_i}$ is a Euclidean metric sparsifier for $\calD$ of quality at most $k^{2 \eta_0}$.
\end{corollary}

\begin{proof}
    The first bound is the proof of \cref{lemma:lipschitz-constants}, and \cref{thm:euclidean-equivalence} gives $\mathsf Q_k = \lipext_k^2$.
    The proof of \cref{thm:euclidean-equivalence} in \cref{app:equivalence} shows that $G^{U_i}$ attains quality $\lipext(\calD, U_i)^2 \leq \lipext_k^2$.
\end{proof}

\begin{proof}[Proof of \cref{thm:euclidean,thm:one-strongly-euclidean}]
    Discard empty sets among the $U_i$.
    \cref{lemma:lipschitz-constants} gives $\lipext_{\max} = O(k^{\eta_0})$, respectively $O(k^{\eta_1})$, and \cref{thm:extendability} gives the distortion bound.
\end{proof}

\section{Lower Bound for Euclidean Sets}
\label{sec:euclidean-lower-bound}

Let $H \ne \set{0}$ be a linear subspace of a finite-dimensional normed space $W$ that inherits the norm~of~$W$.
The \emph{relative projection constant} $\lambda(H, W)$ denotes the least norm of a linear projection $P : W \to H$ that fixes $H$.
Formally, this is
\begin{equation*}
    \lambda(H, W) = \inf \set{\norm{P}_{W \to H} \mid P : W \to H \text{ linear}, P|_H = I_H},
\end{equation*}
where $\norm{P}_{W \to H} = \sup_{w \in W, w \not= 0} {\norm{Pw}} / {\norm{w}}$. \cref{thm:lipschitz-linear} gives a classical connection between linear extensions and Lipschitz extensions.

\begin{theorem}[\citep{lindenstrauss1964nonlinear}]
    \label{thm:lipschitz-linear}
    For finite-dimensional $H \subset W$, we have
    \begin{equation*}
        \inf_{F : W \to H,\ F|_H = I_H} \Lip(F) = \lambda(H, W).
    \end{equation*}
\end{theorem}

\subsection{The cycle space of a graph}

Let $G = (V, E)$ be a connected graph with $n$ vertices, $m$ edges and girth $g \ge 3$.
Fix an arbitrary orientation of the edges, so that each edge $e = (u, v)$ goes from $u$ to $v$.
We view $x \in \RR^E$ as a flow that sends $x_e$ units along edge $e$.
When $x_e < 0$, the flow goes backwards along the edge.
We identify each edge with its indicator vector in $\RR^E$.
The net inflow at a vertex $w$ is $(\partial x)_w = \sum_{e \in \delta^-(w)} x_e - \sum_{e \in \delta^+(w)} x_e$.
Moreover, the net inflow is a linear map $\partial : \RR^E \to \RR^V$.
The cycle space $Z = \ker \partial \subset \RR^E$ consists of the flows that are conserved at every vertex.
For instance, it contains every cycle, which is $+1$ on the edges the cycle traverses along their orientation, $-1$ on those it traverses against it, and $0$ elsewhere.
The dimension of the cycle space is $r = m - n + 1$, and we assume $r > 0$.

Next, choose a root $o$ and a spanning tree of $G$ of depth $D$.
Let $p_v \in \RR^E$ be the signed indicator of the tree path from $o$ to $v$, that is, the unit flow from $o$ to $v$ along this path, so that $\partial p_v = \mathds{1}_v - \mathds{1}_o$, where $\mathds{1}_w \in \RR^V$ is the indicator of a vertex $w$.
For the non-tree edges $e_j = (u_j, v_j)$, let $c_j = p_{u_j} + e_j - p_{v_j}$ for $j \in [r]$.
Notice that $\set{c_j}_{j = 1}^{r}$ are linearly independent.
Indeed, suppose that $\sum_{j = 1}^r a_j c_j = 0$ where the coefficients $a_j$ are not all zero.
Only $c_j$ has 1 at the coordinate of the non-tree edge $e_j$, and all other vectors have 0, and so we must have $a_j = 0$ for each $j \in [r]$.
Therefore, the set $\set{c_j}_{j = 1}^{r}$ forms a basis of~$Z$ because $c_j \in Z$ and $\dim Z = r$.
Let $Ca = \sum_j a_j c_j$.
The vectors $p_v$, $v \neq o$, complement $Z$, since their images $\mathds{1}_v - \mathds{1}_o$ under $\partial$ form a basis of the zero-sum vectors in~$\RR^V$.
Hence, a linear map on $\RR^E$ is determined by its values on $Z$ and on the $p_v$.

\subsection{A norm in which the cycle space is Euclidean}

We define a new norm on $W = \RR^E$ that interpolates between the $\ell_2$ and the $\ell_1$ norm:
\begin{equation}\label{eq:W}
    \norm{x}_W = \inf_{z \in Z} \paren*{\norm{z}_2 + \frac{\norm{x - z}_1}{\sqrt g}}.
\end{equation}
Its unit ball is $B_W = \conv\paren{B_{Z,2} \cup \sqrt g\, B_{\ell_1^E}}$.
In particular, $z = 0$ gives $\norm{e}_W \le 1/\sqrt g$ for every edge.
Hence, $\norm{p_v}_W \le D/\sqrt g$.

Each cycle has at least $g$ edges, so $\norm{z}_1 \geq \sqrt g \norm{z}_2$ for $z \in Z$.
Suppose $z = \sum_i a_i \gamma_i$ with $a_i > 0$ is a decomposition into signed cycles, such that each $\gamma_i$ agrees in sign with $z$ on its support.
Then $\norm{z}_2 \leq \sum_i a_i \sqrt{\abs{\gamma_i}} \leq g^{-1/2} \sum_i a_i \abs{\gamma_i} = g^{-1/2} \norm{z}_1$.
For $x \in Z$, applying this to $x - z$ in \eqref{eq:W} gives $\norm{x}_W \ge \norm{x}_2$, and $z = x$ gives equality.
Thus, $Z$ inherits the Euclidean norm from $W$.

\subsection{The projection constant in the graph construction}

Let $\calP$ be the family of linear maps $P : \RR^E \to Z$ that fix $Z$, so that $\lambda(Z, W) = \min_{P \in \calP} \norm{P}_{W \to Z}$.
The norm of each $P \in \calP$ is determined by the images of the edges.

\begin{lemma}\label{lem:norm}
    For every $P \in \calP$, we have $\norm{P}_{W \to Z} = \max\set{1, \sqrt{g} \max_{e \in E} \norm{Pe}_2}$.
    As a consequence, $\max\set{1, \sqrt{gr/m}} \leq \lambda(Z, W) \leq \sqrt{g}$.
\end{lemma}

\begin{proof}
    Because $Pw \in Z$ for $w \in W$, we have $\norm{Pw}_W = \norm{Pw}_2$, and it suffices to maximize $\norm{Pw}_2$ over the two sets whose convex hull is $B_W$.
    On the ball $B_{Z,2}$ the maximum is one, since $P$ fixes $Z$.
    On the ball $\sqrt g B_{\ell_1^E}$ it is $\sqrt g \max_e \norm{Pe}_2$, attained at a vertex $\pm \sqrt g\, e$.
    Combining these yields the first equality.

    Next, $P$ fixes an orthonormal basis of $Z$.
    Extending it to an orthonormal basis of $\RR^E$ gives $r \leq \norm{P}_{\mathrm{HS}}^2 = \sum_{e \in E} \norm{Pe}_2^2 \leq m \max_e \norm{Pe}_2^2$.
    Together with the first equality, this proves the lower bound.
    For the upper bound, the Euclidean orthogonal projection $Q$ onto~$Z$ satisfies $\norm{Qe}_2 \leq \norm{e}_2 = 1$ for every edge, so $\norm{Q}_{W \to Z} \le \sqrt g$.
\end{proof}

\subsection{From Projection to Lipschitz Extension}

For an integer $N \geq 1$, let $B_N = \set{-N, \ldots, N}^r$.
The cell of a vertex $v$ is $U_{v,N} = \set{p_v + Ca : a \in B_N} \subset W$, and their union is $X_N = \bigsqcup_{v \in V} U_{v,N}$.
The cells are disjoint because $\partial p_v = \mathds{1}_v - \mathds{1}_o$, and they are Euclidean because their differences lie in $Z$.
Let $A_N = U_{o, N}$ and $\phi(Ca) = Ca$.
Let $\lipext(A_N)$ be the optimal Lipschitz constant of an extension of $\phi$ to $X_N$.

\begin{theorem}[From projection constant to Lipschitz constant]\label{thm:transfer}
    Let $\beta_N=2D/(2N-1)$. Then
    \begin{equation}\label{eq:finite}
        \max\set*{1,\frac{\lambda(Z, W)}{1+\beta_N}}
        \le\lipext(A_N)\le\lambda(Z, W).
    \end{equation}
    In particular, $N \ge D+1$ gives $\lipext(A_N) \ge \tfrac12 \sqrt{gr/m}$.
\end{theorem}

\begin{proof}
    Any linear projection $P \in \calP$ is a $\norm{P}_{W \to Z}$-Lipschitz extension of $\phi$.
    Taking the infimum gives the upper bound $\lipext(A_N) \leq \lambda(Z, W)$.

    Now, take an $L$-Lipschitz extension $F : X_N \to Z$ that coincides with $\phi$ on $A_N$.
    For $a \in B_N$, let $h_v(a) = F(p_v + Ca) - F(Ca) = F(p_v + Ca) - Ca$ be the offset of the cell $U_{v,N}$.
    Let $b_v = \abs{J_N}^{-1} \sum_{a \in J_N} h_v(a)$ be average of $h_v$ over the inner box $J_N = \set{-N+1, \ldots, N-1}^r$.
    Since $F(Ca) = Ca$ and $\norm{p_v}_W \le D/\sqrt g$, notice that $h_o = 0$ and $\norm{h_v(a)}_2 \le LD/\sqrt g$.

    For an edge $e = (u, v)$ let $q_e = p_u + e - p_v$.
    It is zero on tree edges and equals $c_j$ on $e_j$.
    When $e$ is a tree edge, let $\eta_e = 0$.
    When $e = e_j$ is a non-tree edge, let $\eta_e \in \ZZ^r$ be the indicator vector with 1 at position $j$.
    This way we have $C \eta_e = q_e$.

    Now because the $p_v$ for $v \ne o$ complement $Z$, we can define the linear map $P$, such that $P|_Z = I_Z$ and $Pp_v = b_v$.
    Then $Pe = q_e + b_v - b_u$ for the edge $e = (u, v)$.
    For $a \in J_N$, the two points $p_u + Ca$ and $p_v + C(a + \eta_e)$ differ by $e$.
    Then,
    \begin{equation*}
        F(p_v + C(a + \eta_e)) - F(p_u + Ca) = h_v(a + \eta_e) - h_u(a) + C \eta_e,
    \end{equation*}
    and
    \begin{align*}
        \norm{F(p_v + C(a + \eta_e)) - F(p_u + Ca)}_2
         & \leq L \norm{p_v + C \eta_e - p_u}_W \\
         & = L \norm{p_v + q_e - p_u}_W \\
         & = L \norm{e}_W.
    \end{align*}
    Therefore, $\norm{q_e + h_v(a + \eta_e) - h_u(a)}_2 \le L/\sqrt g$ for every $a \in J_N$ because $\norm{e}_W \leq {1} / {\sqrt{g}}$.

    Averaging this inequality over $a \in J_N$ for fixed edge $e$, and using the triangle inequality gives $\norm{q_e + \tilde b_v - b_u}_2 \le L/\sqrt g$, where $\tilde b_v = \abs{J_N}^{-1} \sum_{a \in J_N} h_v(a + \eta_e)$ is the average of $h_v$ over the shifted box $J_N + \eta_e$.
    If $e$ is a tree edge, then $\eta_e = 0$ and $\tilde b_v = b_v$.
    Otherwise, $\eta_e$ is the indicator of position $j$, and $J_N + \eta_e$ gains $\set{a \in J_N + \eta_e \mid a_j = N}$ and loses $\set{a \in J_N \mid a_j = -N+1}$.
    All other terms of the two sums cancel out, and each set has $(2N-1)^{r-1} = \abs{J_N}/(2N-1)$ points, so the offset bound gives $\norm{\tilde b_v - b_v}_2 \le \frac{2}{2N-1} \cdot \frac{LD}{\sqrt g} = \beta_N \frac{L}{\sqrt g}$.
    Hence, we get
    \begin{equation*}
        \norm{Pe}_2 = \norm{q_e + b_v - b_u}_2 \le \norm{q_e + \tilde b_v - b_u}_2 + \norm{\tilde b_v - b_v}_2 \le \frac L{\sqrt g}(1+\beta_N).
    \end{equation*}

    We have $L \ge 1$ because $F$ is the identity on $A_N$, so \cref{lem:norm} yields $\lambda(Z, W) \le \norm{P}_{W \to Z} \le (1+\beta_N)L$.
    Taking the infimum over $F$ gives the lower bound.
\end{proof}


\begin{corollary}\label{cor:euclidean-lower-bound}
    For every $k \geq 2$ there is a metric space that is a union of $k$ Euclidean sets with $\lipext_{\max} = \bigomega{\sqrt{\log k}}$.
\end{corollary}

\begin{proof}
    By~\cite{ES63}, there is a $3$-regular graph $G$ with $n \leq k$ vertices and girth $g = \Omega(\log k)$.
    Then $m = 3n / 2$ and $r = m - n + 1 > m / 3$, so $gr / m = \Omega(\log k)$.
    Let $X = X_N$ for $N \geq D + 1$, which is a union of the $n$ Euclidean sets $U_{v, N}$, and adjoin $k - n$ further points, each forming its own set, at distance $\Delta$ from every other point, where $\Delta$ is the diameter of $X_N$.
    Restricting an extension from $X$ to $X_N$ does not increase its Lipschitz constant, so \cref{thm:transfer} gives $\lipext_{\max} \geq \lipext(A_N) \geq \tfrac12 \sqrt{gr / m} = \bigomega{\sqrt{\log k}}$.
\end{proof}

\section{Lower Bound for 1-Strongly Euclidean Sets}
\label{sec:one-strongly-euclidean-lower-bound}

We show that the Lipschitz extension constant can grow even when every set in the union is $1$-strongly Euclidean.
We use a variant of the twisted cube of Naor and Rabani~\cite[Section~4.1]{NR17}, with cross distances chosen so that one layer remains Euclidean after adjoining any one point.

\begin{theorem}\label{thm:one-strongly-euclidean-lower-bound}
    For every integer $k \geq 2$, there is a finite metric space $X = U_1 \cup \cdots \cup U_k$ such that each $U_i$ is $1$-strongly Euclidean in $X$ and
    \[
        \lipext_{\max} \geq \frac{1}{2} (\log_2 k)^{1/4} = \Omega(\sqrt[4]{\log k}).
    \]
\end{theorem}

\subsection{The two-layer cube}

Let $Q_m = \{0,1\}^m$ with Hamming distance $\Ham$, and take two copies $A = \{a_x : x \in Q_m\}$ and $B = \{b_x : x \in Q_m\}$.
For $r > 0$, let $s = \sqrt{r^2 + 1} - r = 1 / (\sqrt{r^2 + 1} + r)$.
On $X_m = A \cup B$, define
\begin{equation}\label{eq:one-strongly-cube-metric}
    \begin{aligned}
        d(a_x, a_y) & = \sqrt{\Ham(x, y)}, \\
        d(a_x, b_y) & = \sqrt{r^2 + \Ham(x, y)}, \\
        d(b_x, b_y) & = \min\{s \Ham(x, y), \sqrt{\Ham(x, y)}\}.
    \end{aligned}
\end{equation}
Thus, $A$ is the usual Euclidean cube, while distances in $B$ are shortened at small Hamming distances.

\begin{lemma}\label{lem:one-strongly-cube-metric}
    For every integer $m \geq 1$ and $r > 0$, \eqref{eq:one-strongly-cube-metric} defines a metric on $X_m$, and $A$ is $1$-strongly Euclidean in $X_m$.
\end{lemma}

\begin{proof}
    For each fixed $y \in Q_m$, the map $a_x \mapsto (x,0)$ and $b_y \mapsto (y,r)$ realizes $A \cup \{b_y\}$ isometrically in $\RR^{m+1}$.
    This proves the one-point Euclidean property and all triangle inequalities with at least two vertices in $A$.

    The function $t \mapsto \min\{s t, \sqrt t\}$ is increasing, concave and vanishes at zero, so it is subadditive and its composition with $\Ham$ is a metric on $B$.
    It remains to check triangles with one vertex in $A$ and two in $B$.

    For every integer $q \geq 0$, concavity of $t \mapsto \sqrt{r^2 + t} - r$ and the definition of $s$ give $\sqrt{r^2 + q} - r \leq \min\{s q, \sqrt q\}$.
    Fix $a_z, b_x, b_y$, and write $p = \Ham(z, x)$, $p' = \Ham(z, y)$ and $q = \Ham(x, y)$.
    Since $\abs{p - p'} \leq q$, subadditivity of the same function gives
    \[
        \abs{d(a_z, b_x) - d(a_z, b_y)} \leq \sqrt{r^2 + \abs{p - p'}} - r \leq \sqrt{r^2 + q} - r \leq d(b_x, b_y).
    \]
    The remaining inequality follows from
    \[
        d(b_x, b_y) \leq \sqrt q \leq \sqrt p + \sqrt{p'} \leq \sqrt{r^2 + p} + \sqrt{r^2 + p'} = d(a_z, b_x) + d(a_z, b_y).
    \]
\end{proof}

\subsection{Extending the cube}

An extension must keep each $b_x$ close to the prescribed image of $a_x$, while neighboring points in $B$ must have nearby images.
Averaging over the cube shows that these two requirements force a large Lipschitz constant.

\begin{lemma}\label{lem:one-strongly-cube-extension}
    For the metric in \eqref{eq:one-strongly-cube-metric}, we have
    \[
        \lipext(A) \geq \frac{\sqrt m}{2 r + s \sqrt m}.
    \]
\end{lemma}

\begin{proof}
    Let $F : X_m \to \ell_2$ be an $L$-Lipschitz extension of the isometry $f(a_x) = x$, and let $u_x = F(b_x)$ and $c = (1/2, \ldots, 1/2) \in \RR^m$.
    If $x^{(j)}$ is obtained from $x$ by flipping coordinate $j$, then $\norm{u_x - x} \leq L r$ and $\norm{u_x - u_{x^{(j)}}} \leq L s$.

    Let $x$ be uniform on $Q_m$.
    Pairing $x$ with $x^{(j)}$ gives, for every $j \in [m]$,
    \[
        \EE[x]{(x_j - 1/2)((u_x)_j - 1/2)} = \frac{1}{2} \EE[x]{(x_j - 1/2)(u_x - u_{x^{(j)}})_j} \leq \frac{L s}{4}.
    \]
    Summing over $j$ and using $\norm{x - c} = \sqrt m / 2$, we obtain
    \[
        \frac{m}{4} = \EE[x]{\ip{x - u_x}{x - c}} + \EE[x]{\ip{u_x - c}{x - c}} \leq \frac{L r \sqrt m}{2} + \frac{m L s}{4}.
    \]
    Rearranging gives $L \geq \sqrt m / (2 r + s \sqrt m)$.
    Since $f$ is an isometry, the same lower bound holds for $\lipext(A)$.
\end{proof}

\subsection{Proof of Theorem~\ref{thm:one-strongly-euclidean-lower-bound}}

\begin{proof}
    Let $m = \ceil{\log_2 k}$ and $r = m^{1/4} / 2$. Then $s \leq 1 / (2 r) = m^{-1/4}$, so \cref{lem:one-strongly-cube-extension} gives
    \[
        \lipext(A) \geq \frac{\sqrt m}{2 r + s \sqrt m} \geq \frac{m^{1/4}}{2} \geq \frac{1}{2} (\log_2 k)^{1/4}.
    \]
    Keep $A$ as one set and partition $B$ into pairs.
    By \cref{lem:one-strongly-cube-metric}, $A$ is $1$-strongly Euclidean in $X_m$.
    Each pair is also $1$-strongly Euclidean, because adjoining one point gives a metric space on at most three points, which is a Euclidean triangle.

    This partition has $1 + 2^{m-1} \leq k$ sets.
    Since $k \leq 2^m$, splitting some pairs into singletons gives exactly $k$ sets while keeping $A$ whole.
    All sets remain $1$-strongly Euclidean, and $\lipext_{\max} \geq \lipext(A)$ proves the theorem.
\end{proof}

\section{Unions of sets with constant distortion}\label{sec:constant-distortion}

We now assume that each $U_i$ embeds into $\ell_2$ with distortion $C_i \geq 1$, so for each $U_i$ there is a map $f_i : U_i \to \ell_2$, such that $d(x, y) \leq \norm{f_i(x) - f_i(y)} \leq C_i d(x, y)$ for every $x, y \in U_i$.
Directly applying \cite[Theorem~1.1]{union16makarychev} for $k = 2$ gives distortion at most $2^{O(k)} \prod_{i = 1}^k C_i$, which is exponential in $k$ even when all $C_i$ are equal to one.

For $c \geq 1$, let $\eta(c)$ be the unique root $\eta \geq c - 1$ of $\paren{ (1 + \eta)^2 - c^2} \paren{ (1 + \eta / 2 )^2 - 2 } = (\pi c^2) / 2$.
For $\eta \geq c - 1$, the left side is at most $0$ while $(1 + \eta / 2)^2 \leq 2$ and afterwards increases to infinity, so the root exists and is unique.
It is increasing in $c$, and $\eta(1) = \eta_0$ by \eqref{eq:improved-eta-zero}.
Moreover, $\eta(c) \leq c - 1 + \pi / c$, because at this value of $\eta$ the two factors on the left are at least $2 \pi$ and $c^2 / 4$.

Next, we establish the following variant of \cref{lem:eliminate}, in which the removed set is only required to embed with distortion $c$.

\begin{lemma}\label{lem:constant-distortion-elimination}
    Let $Y = A \sqcup B$ be a finite pseudometric space with $A \neq \varnothing$, where $B$ embeds into $\ell_2$ with distortion $c \geq 1$.
    For a weighted graph $G$ on $Y$, let $I$ and $J$ be the internal and crossing energies of $B$ as in \cref{lem:eliminate}.
    Then
    \begin{equation}\label{eq:constant-distortion-elimination}
        \calE_{G^A}(d^2) \leq \calE_G(d^2) + (c^2 - 1) I + J + c \sqrt{2 \pi I J}.
    \end{equation}
\end{lemma}

\begin{proof}
    Choose $z_b \in \ell_2$ for $b \in B$ with $d(b, b') \leq \norm{z_b - z_{b'}} \leq c \, d(b, b')$, and let $I_z = \calE_{G[B]}(z) \leq c^2 I$.
    The proof of \cref{lem:eliminate} applies with $I_z$ in place of $I$.
    Indeed, the route $a_p \to b_p \to b_q \to a_q$ still gives $d(a_p, a_q) \leq u_p + \norm{z_{b_p} - z_{b_q}} + u_q$ because $z$ does not contract distances, the middle term is bounded by $2 t I_z$ in \eqref{eq:averaged-distances}, and in the reduction to the case $s_b > 0$ the identity \eqref{eq:minimum-energy} applied to $z$ gives $I_z' \leq I_z$.
    Hence,
    \[
        \calE_{G^A}(d^2) \leq \calE_{G[A]}(d^2) + 2 J + \sqrt{2 \pi I_z J} + I_z \leq \calE_G(d^2) + (c^2 - 1) I + J + c \sqrt{2 \pi I J}.
    \]
\end{proof}

\begin{proof}[Proof of \cref{thm:euclidean-constant-distortion}]
    Fix $i$, keep $U_i$, and partition its complement into at most $k - 1$ subsets of the other sets, as in the proof of \cref{lemma:lipschitz-constants}, so that subsets inherit the distortion bounds.
    For a graph with no edges inside $A$, \cref{lem:constant-distortion-elimination} reads $\calE_{G^A}(d^2) \leq c^2 I + 2 J + c \sqrt{2 \pi I J}$.
    Hence, \cref{lem:improved-asymmetric-extension} applies with $t = 1 + \eta$ and $s = 1 + \eta / 2$ whenever $t \geq c$, $s \geq \sqrt{2}$ and $(t^2 - c^2)(s^2 - 2) \geq \pi c^2 / 2$, that is, whenever $\eta \geq \eta(c)$.
    Embedding with distortion at most $C_{\max}$ is preserved by restriction, so \cref{lem:improved-averaging} with $\eta = \eta(C_{\max}) > 1$ extends every $1$-Lipschitz map on $U_i$ to $X$ with Lipschitz constant at most $k^{\eta(C_{\max})}$.
    Applied to $f_i / C_i$, this extends $f_i$ to $\widetilde f_i : X \to \ell_2$ with $\Lip(\widetilde f_i) \leq C_i \, k^{\eta(C_{\max})}$.
    Since $f_i$ does not contract distances, \cref{lem:terminal-from-extension} gives a $3 \sqrt{2} \, C_{\max} \, k^{\eta(C_{\max})}$-terminal embedding for every~$U_i$, and \cref{thm:composition} gives the claimed distortion.
\end{proof}

\paragraphname{Generative AI Statement} The generalization of Bourgain's Theorem for 2-strongly Euclidean sets (\cref{thm:main}) was obtained by the authors, without any AI usage.
After prompting with the proof of \cref{thm:main}, GPT-6 Astra was able to obtain the presented results for unions of Euclidean (\cref{thm:euclidean}) and 1-strongly Euclidean sets (\cref{thm:one-strongly-euclidean}).
The exposition was significantly rewritten by the authors.
In particular, the authors uncovered the connections to Lipschitz extendability and Euclidean metric sparsifiers.
The authors take full responsibility for the presented results.

\bibliographystyle{alpha}
\bibliography{references}

\newpage\clearpage
\appendix
\section{Proof of the Euclidean sparsifier--extension equivalence}
\label{app:equivalence}

We prove \cref{thm:euclidean-equivalence}.
Fix a partition $U_1,\ldots,U_k$ of $X$, let $\mathcal D$ be the class of pseudometrics on $X$ that are Euclidean on every $U_j$, and set $A=U_i$.
Let $\mathsf Q=\mathsf Q(\mathcal D,A)$.

\paragraph{From sparsifiers to extensions.}
Fix $Q>\mathsf Q$.
Let $d\in\mathcal D$, and let $f:A\to\ell_2$ be $1$-Lipschitz.
The extension claim is immediate when $X$ has one point, so assume $|X|\ge2$.
Since $A$ is finite, we may take $f$ to have values in $\R^m$.

Fix a weighted graph $G$ on $X$, and let $H$ be a Euclidean metric sparsifier for $G$ of quality at most $Q$.
For a scalar map $u:A\to\R$, put $\rho_u(a,b)=|u(a)-u(b)|$.
For every admissible pseudometric $\delta$ extending $\rho_u$, the map $x\mapsto\min_{a\in A}\{u(a)+\delta(x,a)\}$ is a $1$-Lipschitz extension of $u$.
Thus, \eqref{eq:minimum-energy} gives $\calE_{G^A}(u)\le\calE_G(\delta^2)$.
Taking the infimum over $\delta$ and using the reverse inequality from
\cref{lem:canonical}, we get
\[
    \calE_{G^A}(u)=\minext_{\mathcal D}(\rho_u,G)\le\calE_H(u).
\]
Apply this inequality to each coordinate of $f$ and add. We obtain
$\calE_{G^A}(f)\le\calE_H(f)$, and hence
\begin{align*}
    \min_{F|_A=f}\calE_G(F)
     & =\calE_{G^A}(f)\le\calE_H(f) \\
     & \le\calE_H(d^2|_{A\times A}) \\
     & \le Q\minext_{\mathcal D}(d|_{A\times A},G)
    \le Q\calE_G(d^2).
\end{align*}
Here we used that $f$ is $1$-Lipschitz and the weights of $H$ are
nonnegative.

This energy bound holds for every choice of nonnegative edge weights.
Let $\mathcal K=\conv(f(A))$, and let $\mathcal F$ be the compact convex set of extensions $F:X\to\mathcal K$ of $f$.
Projection onto $\mathcal K$ fixes $f(A)$ and cannot increase distances, so the bound also holds with $F\in\mathcal F$.
Let $\mathcal W$ be the probability simplex on unordered pairs of
distinct points of $X$, and put
\[
    \Phi(F,w)=\sum_{x<y}w_{xy}
    \bigl(\|F(x)-F(y)\|^2-Qd(x,y)^2\bigr).
\]
The function $\Phi$ is convex in $F$ and affine in $w$.
For fixed $w$, the energy bound with $g_{xy}=w_{xy}$ gives $\min_{F\in\mathcal F}\Phi(F,w)\le0$.
Sion's minimax theorem therefore gives
\[
    \min_{F\in\mathcal F}\max_{w\in\mathcal W}\Phi(F,w)
    =\max_{w\in\mathcal W}\min_{F\in\mathcal F}\Phi(F,w)\le0.
\]
The maximum over $w$ is the largest of the pairwise differences
$\|F(x)-F(y)\|^2-Qd(x,y)^2$. Thus, some $F\in\mathcal F$ satisfies
\[
    \|F(x)-F(y)\|^2\le Qd(x,y)^2
    \qquad(x,y\in X),
\]
including pairs with $d(x,y)=0$.
Hence, $\lipext(\mathcal D,A)\le\sqrt Q$.
Letting $Q\downarrow\mathsf Q$ gives
\[
    \lipext(\mathcal D,A)^2\le\mathsf Q.
\]

\paragraph{From extensions to sparsifiers.}
Fix $\Lambda>\lipext(\mathcal D,A)$ and a weighted graph $G$ on $X$.
We show that $H=G^A$ is a Euclidean metric sparsifier of quality $\Lambda^2$.

Let $\rho$ be a Euclidean pseudometric on $A$, and choose a Euclidean realization $f:A\to\ell_2$.
For every admissible pseudometric $d$ extending $\rho$, the map $f$ is $1$-Lipschitz and therefore has an extension to $(X,d)$ with Lipschitz constant at most $\Lambda$.
By \eqref{eq:minimum-energy},
\[
    \calE_{G^A}(\rho^2)
    =\min_{F|_A=f}\calE_G(F)
    \le\Lambda^2\calE_G(d^2).
\]
Taking the infimum over $d$ gives the left inequality in
\eqref{eq:sparsifier}; the right inequality follows from
\cref{lem:canonical}.
Thus, $\mathsf Q\le\Lambda^2$.
Letting $\Lambda\downarrow\lipext(\mathcal D,A)$ gives
\[
    \mathsf Q\le\lipext(\mathcal D,A)^2.
\]
Together with the first inequality, this proves the first part of
\cref{thm:euclidean-equivalence}.
Taking suprema over partitions gives $\mathsf Q_k=\lipext_k^2$.

\section{Explicit bounds for a union of two sets}
\label{app:two-pieces}

For $k = 2$ we replace the composition of \cref{thm:composition} by a direct sum of the two extensions, which avoids the unspecified absolute constants of \cref{cor:collapse} and gives explicit distortion bounds.
The direct sum is the same as in \cite[Theorem~2.7]{union16makarychev} and the improvement over their bound of $8.93$ comes from the extension step, where \cref{lem:improved-asymmetric-extension} replaces the extension through an $\alpha$-cover of \cite[Lemma~2.6]{union16makarychev}.

\begin{lemma}\label{lem:two-piece-gluing}
    Let $X = A \sqcup B$ be a finite pseudometric space with $A, B \neq \varnothing$, let $f_A : A \to \ell_2$ and $f_B : B \to \ell_2$ be isometric embeddings, and let $t, s \geq 1$ and $0 < v \leq (s + 1)^2$.
    Suppose that $f_A$ extends to $F_A : X \to \ell_2$ with $\norm{F_A(b) - F_A(b')} \leq t \, d(b, b')$ and $\norm{F_A(a) - F_A(b)} \leq s \, d(a, b)$ for $a \in A$ and $b, b' \in B$, and that $f_B$ extends to $F_B : X \to \ell_2$ with the same factors on pairs inside $A$ and on crossing pairs.
    Then
    \[
        c_2(X)^2 \leq \max\{1 + t^2 + v,\ 2 s^2 + 2 v\} \cdot \frac{(s + 1)^2 + v}{2 v}.
    \]
\end{lemma}

\begin{proof}
    Define $F = F_A \oplus_2 F_B \oplus_2 \sqrt{v} \, d(\cdot, A) \oplus_2 \sqrt{v} \, d(\cdot, B)$.
    For $a, a' \in A$, the first summand is an isometry, the second has factor $t$, the third vanishes, and $\abs{d(a, B) - d(a', B)} \leq d(a, a')$, so $d(a, a')^2 \leq \norm{F(a) - F(a')}^2 \leq (1 + t^2 + v) \, d(a, a')^2$, while pairs inside $B$ are symmetric.
    For $a \in A$ and $b \in B$, the first two summands have factor $s$ and $d(b, A), d(a, B) \leq d(a, b)$, so $\norm{F(a) - F(b)}^2 \leq (2 s^2 + 2 v) \, d(a, b)^2$.

    For the lower bound on crossing pairs, put $d = d(a, b)$, $r_A = d(b, A)$, $r_B = d(a, B)$ and $C = s + 1$, and let $a' \in A$ satisfy $d(a', b) = r_A$.
    Then
    \[
        d \leq d(a, a') + r_A = \norm{F_A(a) - F_A(a')} + r_A \leq \norm{F_A(a) - F_A(b)} + s r_A + r_A,
    \]
    so $\norm{F_A(a) - F_A(b)} \geq (d - C r_A)_+$, and symmetrically $\norm{F_B(a) - F_B(b)} \geq (d - C r_B)_+$.
    Over $r \geq 0$, the function $(d - C r)_+^2 + v r^2$ is minimized at $r = C d / (C^2 + v)$ with value $v d^2 / (C^2 + v)$, hence
    \[
        \norm{F(a) - F(b)}^2 \geq (d - C r_A)_+^2 + v r_A^2 + (d - C r_B)_+^2 + v r_B^2 \geq \frac{2 v}{C^2 + v} \, d^2.
    \]
    Since $v \leq C^2$, this factor is at most one, so $F$ contracts no distance by more than $\sqrt{2 v / (C^2 + v)}$, and the ratio of the squared expansion and contraction bounds proves the claim.
\end{proof}

\begin{theorem}\label{thm:two-piece-bounds}
    Let $(X, d)$ be a metric space with $X = U_1 \cup U_2$.
    If $U_1$ and $U_2$ are Euclidean in $X$, then $c_2(X) < 3.997$.
    If $U_1$ and $U_2$ are $1$-strongly Euclidean in $X$, then $c_2(X) \leq 3.86$.
\end{theorem}

Together with the examples of \cite[Theorem~1.2]{union16makarychev}, which require distortion arbitrarily close to $3$, the first bound gives $3 \leq D_2 < 3.997$.

\begin{proof}
    We may pass to a partition $X = A \sqcup B$, since both conditions are preserved by restriction.
    If one part is empty, $X$ is Euclidean.
    Fix isometric embeddings $f_A$ and $f_B$.

    Suppose first that $A$ and $B$ are Euclidean.
    For a graph on $X$ with no edges inside $A$, \eqref{eq:eliminate} reads $\calE_{G^A}(d^2) \leq I + 2 J + \sqrt{2 \pi I J}$, so by \eqref{eq:improved-h0-tradeoff}, \cref{lem:improved-asymmetric-extension} extends $f_A$ with factor $t$ on $B$ and $s$ on crossing pairs whenever $s \geq \sqrt 2$ and $(t^2 - 1)(s^2 - 2) \geq \pi / 2$, and exchanging the roles of $A$ and $B$ does the same for $f_B$.
    Take $s = {187} / {125}$, $v = {413} / {100}$ and $t^2 = 2 s^2 + v - 1 = 7.606032$.
    Then $(t^2 - 1)(s^2 - 2) = 1.572341312512 > 11 / 7 > \pi / 2$ and $v \leq (s + 1)^2$, and the two terms in the maximum of \cref{lem:two-piece-gluing} coincide, so
    \[
        c_2(X)^2 \leq \frac{(s^2 + v) \paren*{(s + 1)^2 + v}}{v} < 15.975 < 3.997^2.
    \]

    Suppose now that $A$ and $B$ are $1$-strongly Euclidean, so $B \cup \{a\}$ is Euclidean for every $a \in A$, and symmetrically.
    For a graph with no edges inside $A$, \cref{lem:improved-one-point-energy} gives
    \[
        \calE_{G^A}(d^2) \leq \frac{1}{2} \paren*{\sqrt{J} + \sqrt{J + I + \sqrt{I (I + J)}}}^2,
    \]
    and the substitution $z = \sqrt{1 - \sqrt{I / (I + J)}}$ preceding \eqref{eq:improved-one-point-gap} shows that this is at most $t^2 I + s^2 J$ for all $I, J \geq 0$ precisely when
    \[
        D(z) = t^2 (1 - z^2)^2 + s^2 z^2 (2 - z^2) - \frac{1}{2} (z + 1)^2 (2 - z^2) \geq 0 \qquad (0 \leq z \leq 1).
    \]
    Take $s = 143 / 100$, $t^2 = 3 s^2 + s - 1 = 6.5647$ and $v = s (s + 1)$.
    Then
    \begin{align*}
        D(z) & = 5.5647 - 2 z - 9.5396 z^2 + z^3 + 5.0198 z^4 \\
        D'(z) & = -2 - 19.0792 z + 3 z^2 + 20.0792 z^3.
    \end{align*}
    On $[0, 9 / 10]$, the bounds $z^2 \leq 9 z / 10$ and $z^3 \leq 81 z / 100$ give $D'(z) \leq -2 - 0.115048 z < 0$, so $D$ is decreasing there.
    On $[9 / 10, 1]$, the increasing function $D''(z) = -19.0792 + 6 z + 60.2376 z^2$ is at least $D''(9 / 10) > 35$, while exact substitution at $z_0 = 191 / 200$ gives $D(z_0) > 7.476 \cdot 10^{-4}$ and $0 < D'(z_0) < 4.1 \cdot 10^{-3}$, so
    \[
        D(z) \geq D(z_0) + D'(z_0) (z - z_0) + \frac{35}{2} (z - z_0)^2 \geq D(z_0) - \frac{D'(z_0)^2}{70} > 0.
    \]
    Hence, $D \geq 0$ on $[0, 1]$, and \cref{lem:improved-asymmetric-extension} provides both extensions with these factors.
    Since $v \leq (s + 1)^2$ and $1 + t^2 + v = 2 s^2 + 2 v = 4 s^2 + 2 s$, \cref{lem:two-piece-gluing} gives
    \[
        c_2(X)^2 \leq (4 s^2 + 2 s) \cdot \frac{(s + 1)(2 s + 1)}{2 s (s + 1)} = (2 s + 1)^2,
    \]
    that is, $c_2(X) \leq 2 s + 1 = 3.86$.
\end{proof}

\section{Reduction to finite metric spaces}
\label{app:finite-reduction}

For completeness, we include the reduction from separable to finite metric spaces used in \cref{sec:two-strongly-euclidean}.
The argument follows \cite[Lemma~2.9, part~I]{union16makarychev}.

\begin{lemma}[Finite subsets determine distortion]\label{lem:finite-subsets}
    Let $(X, d)$ be a separable metric space, and let $D \geq 1$.
    If every finite subset of $X$ embeds into $\ell_2$ with distortion at most $D$, then so does $X$.
\end{lemma}

\begin{proof}
    Choose a countable dense set $\{x_0, x_1, \ldots\}$ in $X$.
    For every $N$, the finite set $\{x_0, \ldots, x_N\}$ has an embedding $F_N$ with distortion at most $D$, normalized by $F_N(x_0) = 0$ and $d(x, y) \leq \norm{F_N(x) - F_N(y)} \leq D \, d(x, y)$.

    The inner products $\ip{F_N(x_m)}{F_N(x_n)}$ are bounded for every fixed pair, so a diagonal subsequence makes all of them converge.
    Every finite matrix of limiting inner products is positive semidefinite, so the limits are realized by vectors $F(x_n)$ in a separable Hilbert space, and the distance bounds persist.
    Completeness extends $F$ from the dense set to all of $X$ with the same bounds.
\end{proof}

Let $X = U_1 \cup \cdots \cup U_k$ be as in any of \cref{thm:main,thm:euclidean,thm:one-strongly-euclidean,thm:extendability,thm:euclidean-constant-distortion}, and let $S \subseteq X$ be finite.
The nonempty sets $S \cap U_i$ cover $S$, and they inherit each property we impose on the $U_i$: being Euclidean or $h$-strongly Euclidean in $X$, and embedding into $\ell_2$ with a given distortion, are preserved by restriction.
So is the bound $\lipext(U_i) \leq \Lambda$.
By Kirszbraun's Theorem~\cite{kirszbraun1934zusammenziehende}, a Lipschitz map on $S \cap U_i$ extends to the Euclidean set $U_i$ with the same Lipschitz constant, then to $X$ with constant multiplied by $\Lambda$, and restricting the result to $S$ does not increase it.
Moreover, $X$ is separable, since each $U_i$ embeds into $\ell_2$ with finite distortion.
Hence, by \cref{lem:finite-subsets}, each upper bound on $c_2(X)$ in these theorems follows from its finite case.

\end{document}